\documentclass[10pt, twocolumn, comsoc]{IEEEtran}

\usepackage{pgfplots}
\pgfplotsset{compat=1.18}
\usepackage{filecontents}

\usepackage{booktabs}
\usepackage{multirow}
\usepackage{xcolor}

\def\argmin{\mathop{\rm arg\,min}}
\usepackage{graphicx,epsfig}
\usepackage[noadjust]{cite}
\usepackage{mcite}
\usepackage{amsfonts,helvet}
\usepackage{fancyhdr}
\usepackage{threeparttable}
\usepackage{epsf,epsfig}
\usepackage{amsthm}
\usepackage{amsmath}
\usepackage{siunitx}
\usepackage{amssymb}
\usepackage{dsfont}
\usepackage{subfigure}
\usepackage{color}
\usepackage[justification=centering]{caption}
\usepackage[linesnumbered,ruled,noend]{algorithm2e}
\usepackage{algpseudocode}
\usepackage{subcaption}
\usepackage{algcompatible}
\usepackage{enumerate}
\usepackage{gensymb}
\usepackage{cancel}
\usepackage{graphicx}
\usepackage{wrapfig}
\usepackage{ragged2e}

\usepackage{bbm}
\usepackage{eucal}

\usepackage{dsfont}
\usepackage{boldline}

\def\ba{{\bf a}}

\def\be{{\bf e}}
\def\bff{{\bf f}}
\def\bg{{\bf g}}
\def\bh{{\bf h}}

\def\bn{{\bf n}}

\def\bp{{\bf p}}

\def\bu{{\bf u}}
\def\bv{{\bf v}}

\def\bx{{\bf x}}
\def\by{{\bf y}}
\def\bz{{\bf z}}

\def\bA{{\bf A}}

\def\bF{{\bf F}}
\def\bG{{\bf G}}
\def\bH{{\bf H}}
\def\bI{{\bf I}}

\def\bK{{\bf K}}

\def\bP{{\bf P}}
\def\bQ{{\bf Q}}
\def\bR{{\bf R}}
\def\bS{{\bf S}}

\def\cA{\mbox{$\mathcal{A}$}}

\def\cK{\mbox{$\mathcal{K}$}}
\def\cL{\mbox{$\mathcal{L}$}}

\def\cN{\mbox{$\mathcal{N}$}}

\def\bbE{\mbox{$\mathbb{E}$}}

\def\bbH{\mbox{$\mathbb{H}$}}

\def\bbR{\mbox{$\mathbb{R}$}}

\def\bTheta{\mbox{$\boldsymbol{\Theta}$}}
\def\bOmega{\mbox{$\boldsymbol{\Omega}$}}

\begin{document}

\title{Neural Kalman Filtering for Unknown Dynamics: Task-Aware Learning with a Koopman Backbone}

\author{Mintaek~Oh,  Jeonghun~Park, Nir Shlezinger, Yonina C. Eldar, and Jinseok~Choi

\thanks{
Mintaek Oh and Jinseok Choi are with the School of Electrical Engineering, Korea Advanced Institute of Science and Technology (KAIST), Daejeon 34141, South Korea (e-mail: ohmin@kaist.ac.kr; jinseok@kaist.ac.kr). 
\\
\indent Jeonghun Park is with the School of Electrical and Electronic Engineering, Yonsei University, Seoul 03722, South Korea (e-mail: jhpark@yonsei.ac.kr).
\\
\indent {Nir Shlezinger is with the School of ECE, Ben-Gurion University of the Negev, Beer-Sheva 8410501, Israel (e-mail: nirshl@bgu.ac.il).}
\\
\indent {Yonina C. Eldar is with Weizmann Institute of Science, Rehovot 7610001, Israel, and also with the Department of Electrical and Computer Engineering, Northeastern University, Boston, MA 02115, USA (e-mail: yonina.eldar@weizmann.ac.il).}
}

}

\maketitle \setcounter{page}{1} 

\begin{abstract}
Recent years have witnessed a growing interest in AI-aided Kalman filters. 
While emerging methodologies, such as KalmanNet, were shown to facilitate tracking in partially known state-space models, they are not directly applicable when the underlying dynamics is unknown.
To overcome this limitation, we extend the KalmanNet philosophy to the unknown-dynamics regime by developing blind Kalman filtering frameworks that learn both the predictor and the correction gain from data, assuming that the state-evolution function and the noise statistics are both unavailable.
To this end, we first introduce a task-aware neural Kalman filtering framework, Blind-KalmanNet, which   carries the learning principle of the Kalman gain into the prediction step through a two-head neural architecture that jointly learns a state-dependent linear surrogate and the Kalman gain from data.
Building on this formulation, we then develop our main framework, Koopman-aided Blind-KalmanNet, which incorporates Koopman operator theory to lift the unknown dynamics into a latent space where the state evolves linearly.
The lifted linear predictor integrates seamlessly into the Blind-KalmanNet structure: the pre-trained deep Koopman network serves as a globally structured predictor, augmented by the task-aware residual surrogate and the learned Kalman gain inherited from Blind-KalmanNet.
Extensive experiments demonstrate that the proposed frameworks achieve competitive performance against baselines, with Koopman-aided Blind-KalmanNet attaining the best accuracy across all considered settings.
\end{abstract}

\begin{IEEEkeywords}
Kalman filter, state estimation, unknown dynamics, KalmanNet, Koopman operator.
\end{IEEEkeywords}

\section{Introduction}

Sequential state estimation from noisy observations is a core task in signal
processing and control, with broad applications including target tracking,
robotics, navigation, wireless localization, and autonomous systems
\cite{hightower2002location, you2020data,
talbot2025continuous}. 
A common mathematical abstraction is the state-space (SS) model, in which a
hidden state evolves over time according to a possibly nonlinear state-evolution
function and is inferred from noisy observations.
This abstraction encompasses a broad range of practical systems, making
sequential state estimation a fundamental and widely studied problem.

In many practical scenarios, however, the state-evolution function is nonlinear
or too complex to be reliably specified, while the associated noise statistics
are also unknown.
Model-based filters can suffer substantial performance degradation under such
model mismatch, motivating sequential state estimation frameworks that do not
require an explicitly specified state-evolution model or noise statistics.

\subsection{Related Works}

We first review model-based approaches that assume an explicitly specified SS model.
When the dynamics and observation models are known and linear with Gaussian noise, the Kalman filter (KF)~\cite{kalman1960new} provides a computationally efficient recursive minimum mean-squared error (MMSE) estimator.
For nonlinear systems, the extended KF (EKF)~\cite{anderson1979optimal} propagates the first- and second-order moments through a local linearization of the dynamics, while the unscented KF~\cite{julier2004unscented} propagates a deterministic set of sigma points to capture the posterior moments without Jacobians. 
Beyond the Gaussian regime, the particle filter~\cite{gordon1993novel} approximates the full posterior via weighted Monte Carlo samples, accommodating arbitrary nonlinear and non-Gaussian models at a higher computational cost.

To handle strong nonlinearities while retaining a linear recursion, a parallel
line of work combines the KF with Koopman operator theory
\cite{koopman1931hamiltonian, mezic2005spectral, williams2015data},
which represents nonlinear dynamics as a linear evolution on a lifted space of
observables.
Early Koopman-based observer frameworks \cite{surana2016linear, surana2016koopman} exploit such a lifted linear representation to enable Kalman-like state estimation for nonlinear systems.
In practice, however, a suitable finite-dimensional Koopman representation is generally not known a priori and must be identified by approximating the lifting functions and the associated linear operator.
This has motivated data-driven Koopman approaches.
For example, the robust data-driven Koopman KF \cite{netto2018robust} estimates a finite-dimensional Koopman representation from trajectory data and performs state estimation in the resulting lifted linear model.
More recently, deep Koopman networks (DKNs)
\cite{takeishi2017learning, lusch2018deep, shi2022deep} use a deep neural network (DNN) to learn the lifting functions together with the associated Koopman operator from data, and the deep Koopman KF \cite{sui2024deep} integrates such a learned representation with Kalman filtering for process denoising.
Despite these advances, existing Koopman-based filtering approaches generally retain an analytical Kalman update in the lifted model and therefore require noise covariance information that can be difficult to specify in practice.

A complementary line of model-based work relaxes the requirement of a fully specified model.
A long line of adaptive KF (AKF) estimates the unknown noise covariances online while preserving the standard Kalman recursion~\cite{mehra1970identification, myers1976adaptive, sarkka2009recursive}; these methods, however, still assume that the state-evolution function is fully known and only adapt the noise statistics.
More generally, expectation-maximization (EM)-based approaches have been used to estimate unknown SS-model parameters by alternating between state estimation or smoothing and model-parameter updates~\cite{shumway1982approach, gannot2008kalman}.
Along this line,
blind KF (BKF)~\cite{sharma2020blind} jointly estimates the state together with the unknown linear state and observation matrices via EM over sliding mini-batches.
However, BKF estimates only a single time-invariant linear pair per window under a linear-Gaussian SS model with manually specified noise covariances, which limits its applicability to strongly nonlinear dynamics.
Overall, while these model-based filters offer interpretability and structure, they either require explicit knowledge of the dynamics or are confined to restrictive linear models.



To overcome these limitations, recent advances have explored diverse deep learning paradigms for sequential state estimation. 
One prominent data-driven direction aims to enable the estimation directly from observations without relying on an explicit state-evolution function.
\emph{DANSE}~\cite{ghosh2024danse} represents a key approach along this line. 
It employs a DNN to parameterize a Gaussian predictive state distribution directly from past measurements, thereby bypassing an explicit state-evolution model.
Under a linear-Gaussian observation model with known noise covariance, the posterior update is obtained analytically, which also enables training without state labels through the predictive observation likelihood.
The framework has been extended to semi-supervised estimation from compressed measurements~\cite{ghosh2026semi}, while \emph{pDANSE}~\cite{ghosh2025pdanse} handles nonlinear observation models using particle-based likelihood weighting.
The DANSE family directly parameterizes the predictive prior from the measurement history rather than propagating the previous posterior through a state-transition model.
It requires the observation noise statistics to be known and assumes a Gaussian predictive prior, while pDANSE additionally incurs Monte-Carlo sampling at each time step.

In contrast, another major paradigm within model-based deep learning~\cite{shlezinger2023model} retains the recursive structure of KF while learning the missing components from data.
A representative approach in this category is \emph{KalmanNet}~\cite{revach2022kalmannet}, which targets the partially known dynamics regime by preserving the Kalman recursion while replacing the analytical Kalman gain with a recurrent neural network (RNN)~\cite{lipton2015critical} learned from data. 
This makes KalmanNet robust under mismatch in the state-evolution function and, being learned end-to-end, eliminates the need for noise statistics.
Building on this philosophy, numerous variants have been proposed along complementary axes.
One line modifies the gain-computing architecture: \emph{Split-KalmanNet}~\cite{choi2023split} decomposes the gain computation into two DNN modules that separately learn the prior-state and innovation covariance surrogates; \emph{Bayesian-KalmanNet}~\cite{dahan2025bayesian} employs Bayesian networks to quantify uncertainty; and \emph{KalmanFormer}~\cite{shen2025kalmanformer} and \emph{GSP-KalmanNet}~\cite{buchnik2024gsp} replace the recurrent module with attention- and graph-based architectures, respectively.
A second line relaxes a different component of the required domain knowledge: \emph{Latent-KalmanNet}~\cite{buchnik2023latent} targets tracking from high-dimensional measurements whose observation function is unknown, by learning an encoder that maps the observations into a latent representation in which the observation model reduces to a known selection matrix, and applying KalmanNet in that latent space.
A third line improves adaptation efficiency: \emph{MAML-KalmanNet}~\cite{chen2025maml} employs meta-learning to reduce the amount of labeled data and the number of training rounds needed for a new setting, whereas \emph{Adaptive-KalmanNet}~\cite{ni2024adaptive} uses a hypernetwork to accommodate changes in the SS model without retraining.
A broader overview of this rapidly growing family is provided in~\cite{shlezinger2025artificial}.
Nonetheless, these methods fundamentally rely on domain knowledge, or a sufficiently accurate surrogate, to form the prediction, and thus cannot be applied when the state-evolution function is entirely unavailable.

Taken together, these lines of work do not simultaneously provide a Kalman-structured estimator that operates without an explicit state-evolution function and noise statistics.
Motivated by this gap, we extend the KalmanNet philosophy to the unknown-dynamics setting by learning both the predictor and the correction gain from data while preserving the Kalman predictor--corrector structure.

\subsection{Contributions}
In this paper, we propose task-aware neural Kalman filtering for sequential state estimation under unknown state-evolution functions and noise statistics.
To this end, we first retain the EKF prediction flow and replace its local linearization with a learned surrogate, and then introduce a Koopman lifting mapping that represents the unknown dynamics by a globally shared linear operator.
Our main contributions are summarized as follows:
\begin{itemize}
    \item 
    We formulate sequential state estimation under unknown dynamics, where neither the state-evolution function nor the noise statistics are available, and address it by extending the KalmanNet philosophy: 
   both the prediction and correction steps are learned in a task-aware manner. This principle is conceptually akin to discriminative learning \cite{10243463}, in that the learned components are optimized directly for the state-estimation task rather than for identifying the underlying model.
    
    \item
    We first introduce Blind-KalmanNet, which realizes this principle through a two-head architecture in which one recurrent network generates a state-dependent linear surrogate of the unknown dynamics from the filtering history while the other produces the Kalman gain for observation correction. The learned surrogate serves as a data-driven, task-aware counterpart to the local linearization of the EKF, and we establish its trainability by deriving the loss gradient with respect to the learned transition matrix, enabling joint end-to-end training of the two heads.
    
    \item
    Building on this Blind-KalmanNet formulation, we develop our main framework, Koopman-aided Blind-KalmanNet, which lifts the unknown dynamics into a latent space where the state evolves linearly under a globally shared Koopman operator, a representation naturally favorable for Kalman filtering. A two-stage strategy pre-trains a deep Koopman network as a structured prediction backbone, then freezes it while a task-aware residual surrogate and the Kalman gain are jointly learned, thereby retaining the Blind-KalmanNet structure.

    \item 
     Extensive experiments on linear and  nonlinear dynamics with linear and nonlinear observations, as well as a real-world drone localization dataset with non-Gaussian sensing noise, compare the proposed frameworks with model-based and AI-aided baselines tailored for unknown dynamics. Koopman-aided Blind-KalmanNet attains the best accuracy across all considered settings, approaching oracle filters with perfect dynamics knowledge, while Blind-KalmanNet recovers most of this gain with a smaller model and single-stage training.
\end{itemize}

\paragraph*{Notation}
The superscripts $(\cdot)^{\sf T}$ and $(\cdot)^{-1}$ denote the transpose and matrix inversion, respectively. 
We use ${\bf{I}}_N$ for the identity matrix of size $N \times N$ and $\bf 0$  for a zero vector with proper dimension.
We let $\bbR^{m \times n}$ be the $m \times n$ dimensional real space.
We use ${\rm{vec}}(\cdot)$ for vectorization and $\|\cdot\|$ for the Euclidean norm, and $\|\cdot \|_F$ for the Frobenius norm.
For a set $\boldsymbol{\cA}$, $\left| \boldsymbol{\cA} \right|$ indicates cardinality.

\section{System Model and Problem Formulation}

Considering time instance $t \in \CMcal{T}$, where $\CMcal{T} = \{1, 2, \ldots, T\}$, the SS model is given by
\begin{align}
    \label{eq:x_t}
    \bx_{t} &= \bff(\bx_{t-1}) + \be_{t} \in \bbR^{m},
    \\
    \label{eq:y_t}
    \by_{t} &= \bh(\bx_{t}) + \bn_{t} \in \bbR^{n},
\end{align}
where $\bff(\cdot)$ is a state-evolution function, $\bh(\cdot)$ is an observation function, $\be_t$ is the process noise capturing uncertainty in the state evolution, and $\bn_t$ is the observation noise in the measurement.
The process noise $\be_t$ and the observation noise $\bn_t$ are zero-mean, mutually and temporally independent, with covariances $\bQ = \bbE[\be_t\be_t^{\sf T}]$ and $\bR = \bbE[\bn_t\bn_t^{\sf T}]$, respectively.
Throughout this paper, we consider the \emph{unknown-dynamics} scenarios, where the state-evolution function $\bff(\cdot)$ and the noise covariances $\bQ$ and $\bR$ are unknown, whereas the observation function $\bh(\cdot)$ is assumed known since it is typically dictated by sensor geometry and measurement physics and can be obtained via system design or calibration.

The sequential state estimation problem seeks the MMSE estimate of $\bx_t$ given the observations up to time $t$.
Mathematically, the following optimization problem needs to be solved at each time $t$:
 \begin{align}
    \label{eq:main}
     \argmin_{\hat{\bx}_t}& \;\;  \bbE \left[ \left.\left\|\bx_t - \hat{\bx}_t \right\|^2 \right| \, \by_1, \ldots, \by_t \right].
 \end{align}
When the SS model is linear-Gaussian and known, the KF offers an optimal recursive solution to this problem.

To learn the estimator in \eqref{eq:main} without explicit dynamics knowledge,
we assume access during training to a labeled dataset
$\big(\bx_{1:T}^{(\ell)},\by_{1:T}^{(\ell)}\big)$,
$\ell\in\boldsymbol{\CMcal{D}}$, referred to as the ground-truth data (GTD).
The GTD provides only sampled access to the underlying dynamics, while
$\bff(\cdot)$, $\bQ$, and $\bR$ remain unavailable in explicit form.
At test time, only the observations $\by_{1:t}$ are available.
Such labeled trajectories can be collected offline using a high-precision reference system, while only noisy observations are available during deployment.

\section{Preliminaries}

\subsection{Kalman Filtering} \label{subsec:KF}
In this subsection, we recall the conventional KF.
In every step $t$, the KF produces a new estimate $\hat{\bx}_t$ using only the previous estimate $\hat{\bx}_{t-1}$ and the new observation $\by_t$.
Assuming a linear SS model, i.e., $\bff(\bx_t) = \bF \, \bx_t$ and $\bh(\bx_t) = \bH \, \bx_t$, the recursion consists of two primary steps.
In the \textit{prediction} step, the a priori moments of the state and the observation are propagated using the domain knowledge of $\bF$ and $\bH$ as
\begin{align}
    \label{eq:x_prior}
    \hat{\bx}_{t|t-1} &= \bF \, \hat{\bx}_{t-1|t-1},
    &
    \boldsymbol{\Sigma}_{t|t-1} &= \bF \, \boldsymbol{\Sigma}_{t-1|t-1} \, \bF^{\sf T} + \bQ,
    \\
    \label{eq:y_prior}
    \hat{\by}_{t|t-1} &= \bH \, \hat{\bx}_{t|t-1},
    &
    {\bS}_{t|t-1} &= \bH \, \boldsymbol{\Sigma}_{t|t-1} \, \bH^{\sf T} + \bR.
\end{align}
In the \textit{update} step, the a posteriori moments are then obtained:
\begin{align}
     \hat{\bx}_{t|t} = \hat{\bx}_{t|t-1} + \bK_t  \left(\by_t - \hat{\by}_{t|t-1}\right),
    \ 
    \boldsymbol{\Sigma}_{t|t} = \left(\bI - \bK_t \bH \right)  \boldsymbol{\Sigma}_{t|t-1},
\end{align}
where $\by_t - \hat{\by}_{t|t-1}$ is the innovation and $\bK_t$ is the Kalman gain:
\begin{align}
    \label{eq:Kalman_gain}
    \bK_t = \boldsymbol{\Sigma}_{t|t-1} \, \bH^{\sf T} \, \bS^{-1}_{t|t-1}.
\end{align}

The EKF extends the KF for nonlinear $\bff(\cdot)$ and/or $\bh(\cdot)$.
The first-order statistical moments in \eqref{eq:x_prior} and \eqref{eq:y_prior} are replaced with 
\begin{align}
    \hat{\bx}_{t|t-1} &= \bff (\hat{\bx}_{t-1|t-1}),
    \\
    \hat{\by}_{t|t-1} &= \bh(\hat{\bx}_{t|t-1}),
\end{align}
respectively.
The second-order moments are approximated  through the nonlinearity.
The EKF linearizes the differentiable $\bff(\cdot)$ and/or $\bh(\cdot)$ in a time-dependent manner using their Jacobian matrices, evaluated at $\hat{\bx}_{t-1|t-1}$ and $\hat{\bx}_{t|t-1}$:
\begin{align}
    \label{eq:Jacobian}
    {\boldsymbol{\CMcal{F}}}_{t} = \left.\frac{\partial \bff}{\partial \bx}\right|_{\bx = \hat{\bx}_{t-1|t-1}},\;\;\; {\boldsymbol{\CMcal{H}}}_{t} = \left.\frac{\partial \bh}{\partial \bx}\right|_{\bx = \hat{\bx}_{t|t-1}}.
\end{align}
The EKF then replaces $\bF$ and $\bH$ in  \eqref{eq:x_prior}--\eqref{eq:Kalman_gain} with the Jacobian matrices ${\boldsymbol{\CMcal{F}}}_{t}$ and ${\boldsymbol{\CMcal{H}}}_{t}$, respectively.
When the SS model is linear, the EKF coincides with the KF, which achieves the MMSE for linear-Gaussian SS models.

The resulting filter admits an efficient linear recursive structure. 
However, it requires full knowledge of the underlying model and degrades in the presence of model mismatch. 
In addition, when the SS model is strongly nonlinear, the first-order linearization can be inaccurate, leading to degraded EKF performance.
This motivates the augmentation of EKF into KalmanNet  detailed in the next subsection.

\subsection{KalmanNet} \label{subsec:KNet}
The accuracy of SS models impacts the performance of the model-based KFs
significantly.
In scenarios where noise distributions are unknown, we can only use approximated methods to estimate the noise statistics.
This limitation can significantly impact the accuracy of state estimation.
To address this issue, KalmanNet has been proposed as a DNN-aided approach to augment the KF.
This method replaces the analytical computation of the Kalman gain with a DNN that learns it from data. 
We adopt the gated recurrent unit (GRU) \cite{cho2014} following the original KalmanNet \cite{revach2022kalmannet}, whose recurrent structure can implicitly learn the noise statistics from the input features.
This allows it to capture essential information required to
evaluate the Kalman gain with DNN parameters $\boldsymbol{\Theta}$.
Thanks to the properties of KalmanNet, there is no need to explicitly compute any second-order moments, allowing the \textit{update} step of the KF to be replaced by
\begin{align}
    \hat{\bx}_{t|t} = \hat{\bx}_{t|t-1} + \bK_t\!\left(\boldsymbol{\Theta}\right) \left(\by_t -  \hat{\by}_{t|t-1}\right).
\end{align}
The DNN parameters $\boldsymbol{\Theta}$ in the dedicated architecture are trained by optimizing the square error loss function in a supervised fashion.
During the training phase, the empirical mean-squared error (MSE) loss is used to measure the error between the true states $\bx_t$ and their estimated state $\hat{\bx}_{t|t}$.

Substituting the prediction step into the KalmanNet update, the posterior state update can be written as
\begin{align}
    \label{eq:x_pos}
   \hat{\bx}_{t|t} = \bff(\hat{\bx}_{t-1|t-1})  + \bK_t\!\left(\boldsymbol{\Theta}\right) \left(\by_t -  \bh\left(\bff(\hat{\bx}_{t-1|t-1})\right) \right).
\end{align}
We, however, cannot directly apply KalmanNet since we assume $\bff(\cdot)$ is not known.
We address this challenge by proposing task-aware and data-driven approaches.
Before presenting our proposed approaches, we first review two na\"ive data-driven approaches that attempt to operate under unknown dynamics, which serve as baselines.

\section{Na\"ive Approaches and Task-Aware Learning} \label{sec:naive}
In this section, we briefly describe two straightforward approaches that attempt to operate under the unknown-dynamics assumption and will be used for comparison.
The first approach discards the Kalman recursion and trains an end-to-end RNN to directly map the observation sequence to state estimates.
The second approach retains a KalmanNet-style correction step, but replaces $\bff(\cdot)$ with a simple surrogate matrix, while learning the Kalman gain from training data.

\subsection{End-to-End RNN} 
\label{subsec:E2E_RNN}
A direct approach for model-free sequential state estimation is to train an end-to-end RNN that maps the observation sequence without explicitly using the SS model.
Let $\boldsymbol{\mathcal{R}}(\cdot\,;\boldsymbol{\Phi})$ denote an RNN, e.g., based on  GRU~\cite{cho2014} cells, with trainable parameters $\boldsymbol{\Phi}$.
The estimator can be written as
\begin{equation}
    \hat{\mathbf{x}}_t = \boldsymbol{\mathcal{R}}(\mathbf{y}_t, {\bu}_{t-1}; \boldsymbol{\Phi}),
\label{eq:e2e_rnn}
\end{equation}
where ${\bu}_t$ is the hidden state at time step $t$ and $\boldsymbol{\Phi}$ is learned by minimizing the MSE over labeled training trajectories.

While this approach operates without access to either $\bff(\cdot)$ or $\bh(\cdot)$, treating filtering as a generic sequence-to-sequence regression provides only limited inductive bias: the network must implicitly learn the prediction and correction behaviors that are explicitly encoded in Kalman-type recursions, leading to higher sample complexity and sensitivity to distribution shifts such as changes in sequence length, noise statistics, or dynamics regimes.
This limitation can be partially mitigated by imposing more
structure on the recurrent mapping. 
For instance, the recurrent Kalman network \cite{becker2019recurrent} employs two separate recurrent modules for the prediction and update steps within a learned high-dimensional feature space, without relying on SS models.
Such designs nonetheless learn the entire predict--update mapping in a feature space, without explicitly retaining the Kalman recursion.
A complementary strategy, which we examine next, instead preserves the recursion and supplies the missing predictor through a simple surrogate.


\subsection{Regression-based KalmanNet} 
\label{subsec:Reg_KNet}
The end-to-end RNN baseline in Section~\ref{subsec:E2E_RNN} can operate in a fully blind manner, but it discards the predictor--corrector structure of Kalman-type filtering.
As an intermediate approach, we instead retain the KalmanNet update, i.e., we still learn the Kalman gain from data, and only replace the unavailable predictor with a simple surrogate.
Since KalmanNet requires $\bff(\cdot)$ in the prediction step,
we approximate $\bff(\cdot)$ by a linear mapping identified from the GTD via regression.
For each $\ell \in \boldsymbol{\CMcal{D}}$, the GTD state trajectory of length $T$ is
\begin{align}
    \mathbf{X}^{\mathrm{tr}}(\ell)
    = \big[\mathbf{x}^{\mathrm{tr}}_{1}(\ell), \mathbf{x}^{\mathrm{tr}}_{2}(\ell), \ldots,
    \mathbf{x}^{\mathrm{tr}}_{T}(\ell)\big] \in \mathbb{R}^{m \times T}.
\end{align}
From $\mathbf{X}^{\mathrm{tr}}(\ell)$, we form one-step regression pairs by shifting:
\begin{align}
    \mathbf{X}_{-}^{\mathrm{tr}}(\ell)
    &= \big[\mathbf{x}^{\mathrm{tr}}_{1}(\ell), \mathbf{x}^{\mathrm{tr}}_{2}(\ell), \ldots,
    \mathbf{x}^{\mathrm{tr}}_{T-1}(\ell)\big] \in \mathbb{R}^{m \times (T-1)}, \\
    \mathbf{Y}^{\mathrm{tr}}(\ell)
    &= \big[\mathbf{x}^{\mathrm{tr}}_{2}(\ell), \mathbf{x}^{\mathrm{tr}}_{3}(\ell), \ldots,
    \mathbf{x}^{\mathrm{tr}}_{T}(\ell)\big] \in \mathbb{R}^{m \times (T-1)}. 
\end{align}
We then consider a linear surrogate dynamics as
\begin{align}
    \bx_{t} \approx \widehat{\bF} \bx_{t-1},
\end{align}
where $\widehat{\bF}\in \bbR^{m\times m}$ is the state transition matrix to be estimated.

Stacking all training pairs across $\ell \in \boldsymbol{\CMcal{D}}$ yields
\begin{align}
    \mathbf{X} &= \big[\mathbf{X}_{-}^{\mathrm{tr}}(1), \ldots, \mathbf{X}_{-}^{\mathrm{tr}}(|\boldsymbol{\CMcal{D}}|)\big]
    \in \mathbb{R}^{m\times N}, 
    \\
    \mathbf{Y} &= \big[\mathbf{Y}^{\mathrm{tr}}(1), \ldots, \mathbf{Y}^{\mathrm{tr}}(|\boldsymbol{\CMcal{D}}|)\big]
    \in \mathbb{R}^{m\times N},
\end{align}
where $N = |\boldsymbol{\CMcal{D}}|(T-1)$ is the total number of state pairs.
To ensure numerical stability when $\mathbf{X}\mathbf{X}^{\sf T}$ is ill-conditioned, we adopt ridge regression~\cite{hoerl1970ridge, hastie2009elements}:
\begin{align}
    \widehat{\mathbf{F}}
    \!=\! \argmin_{\widehat{\bF}} \|\mathbf{Y}-\widehat{\bF}\mathbf{X}\|_{F}^{2} + \lambda \|\widehat{\bF}\|_{F}^{2} = \big(\mathbf{Y}\mathbf{X}^{\sf T}\big)\big(\mathbf{X}\mathbf{X}^{\sf T} \!+\! \lambda \mathbf{I}\big)^{-1}\!,
\end{align}
where $\lambda>0$ is the regularization strength.

Since KalmanNet has demonstrated robust performance under SS model mismatch 
in various scenarios as shown in \cite{shlezinger2025artificial, revach2022kalmannet, ni2024adaptive}, one might anticipate that a regression-based approach can also achieve strong performance  when $\bff(\cdot)$ is approximated via linear regression.
Consequently, we construct the predictor by plugging the regression-based dynamics into the predictor, i.e., replacing $\bff(\mathbf{x})$ with $\widehat{\mathbf{F}}\mathbf{x}$.
Thus, \eqref{eq:x_pos} becomes
\begin{align}
\hat{\bx}_{t|t}
= \widehat{\bF}\,\hat{\mathbf{x}}_{t-1|t-1} + \mathbf{K}_t(\boldsymbol{\Theta})
\Big(\mathbf{y}_{t} - \bh\big(\widehat{\mathbf{F}}\,\hat{\mathbf{x}}_{t-1|t-1}\big)\Big).
\end{align}
The DNN still learns the Kalman gain $\mathbf{K}_t(\boldsymbol{\Theta})$ from data, while $\widehat{\mathbf{F}}$ is fixed after regression.
We refer to this plug-in baseline as \emph{Regression-based KalmanNet}.
Note that this approach yields a closed-form $\widehat{\bF}$ from the available GTD, which is inherently time-invariant and thus cannot adapt over time.

\subsection{Task-Aware Learning Principle}
\label{subsec:task-aware}

The limitations of the two na\"ive approaches, though different in nature, share a common origin: the learned components are optimized for objectives other than the filtering task in \eqref{eq:main}. The end-to-end RNN abandons the Kalman recursion and must implicitly rediscover both prediction and correction from data, with no structural guarantee that either behavior emerges. 
Regression-based KalmanNet retains the recursion, yet its surrogate $\widehat \bF$ is fitted to a one-step prediction objective and then frozen; the resulting predictor is never informed by the posterior estimation error it ultimately serves, and no subsequent gain learning can compensate for a predictor misaligned with the filtering task.

These observations motivate what we refer to as task-aware learning: every learned component, the predictor as well as the correction gain, is trained jointly under the posterior state-estimation objective in \eqref{eq:main}. 
The surrogate dynamics is thereby shaped not for one-step prediction accuracy, but for its actual role within the predictor–corrector recursion.
This principle generalizes the original KalmanNet philosophy to the unknown-dynamics regime: whereas KalmanNet applies task-aware learning only to the Kalman gain and assumes the predictor is given, we extend it to the predictor itself. 
The following sections develop this principle into the proposed frameworks: Section~\ref{sec:BKNet} first realizes it in its  direct form through a state-dependent linear surrogate, and  Section~\ref{sec:BK2Net} then augments it with a globally structured Koopman predictor.

\section{Blind-KalmanNet} \label{sec:BKNet}
In this section, we introduce Blind-KalmanNet, the first design stage realizing the task-aware learning principle in Section~\ref{subsec:task-aware}: a second DNN parameterizes the unknown state evolution as a state-dependent linear surrogate $\bF_t$, which is trained jointly with the gain network $\bK_t$ under the posterior state-estimation objective. Section~\ref{sec:BK2Net} then builds on this design to develop Koopman-aided Blind-KalmanNet, the main framework of this paper.

\subsection{Blind-KalmanNet Architecture}
\label{subsec:B_KNet}
\begin{figure}[t]    
    {\centerline{\resizebox{1\columnwidth}{!}{\includegraphics{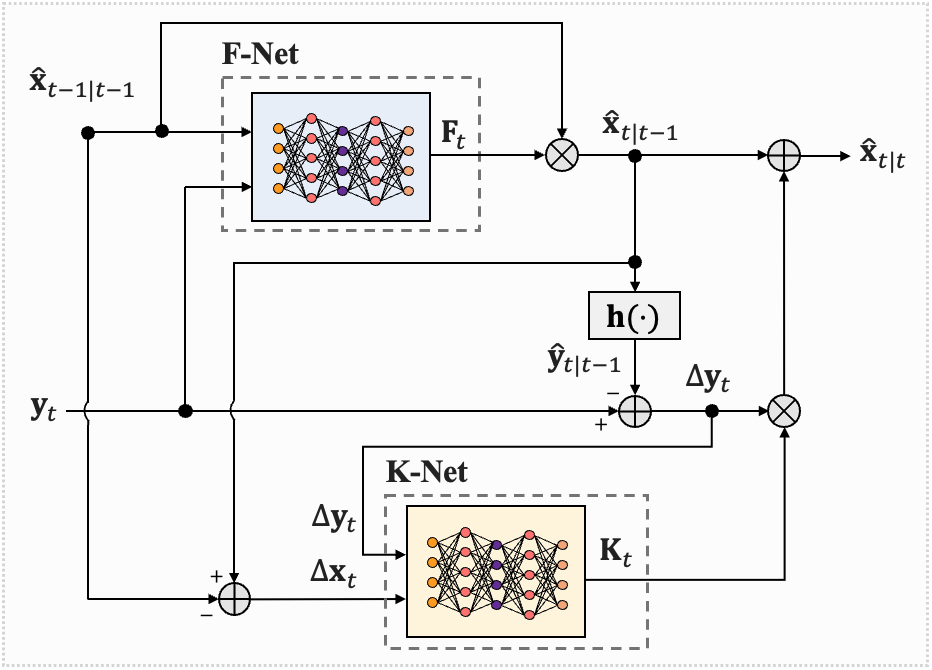}}}
    \caption{Blind-KalmanNet block diagram.}
    \label{fig:B_KNet}} 
\end{figure}
We design {Blind-KalmanNet}, a two-head hybrid filtering architecture tailored for the unknown dynamics setting.
Unlike Regression-based KalmanNet, which uses a fixed surrogate $\widehat{\bF}$ estimated from GTD, Blind-KalmanNet learns a state-dependent surrogate directly under the filtering objective.
While the true state-evolution function $\bff(\cdot)$ can be nonlinear, at each time step $t$, we locally approximate the state transition by a linear mapping:
\begin{align}
    \label{eq:linear_approx}
   \bx_t \approx \bF_t \bx_{t-1}.
    \end{align}
Here, $\bF_t \in \mathbb{R}^{m \times m}$ is a surrogate transition matrix generated from the recent filtering history at each time step.
Whereas the fixed $\widehat{\bF}$ of Regression-based KalmanNet provides a single global linear approximation, $\bF_t$ adapts along the trajectory to the local behavior of $\bff(\cdot)$.
In this sense, F-Net learns to compute the local linearization of the EKF in a task-aware manner without access to $\bff(\cdot)$.
Moreover, since the original KalmanNet has demonstrated robust performance under model mismatch~\cite{revach2022kalmannet}, learning $\mathbf{F}_t$ jointly with the Kalman gain $\mathbf{K}_t$ extends this robustness to the unknown-dynamics setting.

\paragraph{High-level Architecture}
We derive Blind-KalmanNet by isolating the KF computations that rely on unavailable model knowledge.
As depicted in Fig.~\ref{fig:B_KNet}, {Blind-KalmanNet} consists of two DNN modules:
$(i)$ \textbf{F-Net}, which outputs a state transition matrix $\bF_t(\boldsymbol{\Omega}) \in \bbR^{m\times m}$ with its trainable parameters $\boldsymbol{\Omega}$, and $(ii)$ \textbf{K-Net}, which outputs a Kalman gain $\bK_t(\boldsymbol{\Theta}) \in \mathbb{R}^{m\times n}$.
K-Net plays the same role as the gain network of KalmanNet, whereas F-Net is the new component that carries the task-aware treatment into the prediction step.
Given the previous posterior estimate $\hat{\bx}_{t-1|t-1}$ and the new observation $\by_t$, Blind-KalmanNet operates with the Kalman filtering flow:
\begin{align}
    \label{eq:bknet_predict}
    \hat{\bx}_{t|t-1} &= \bF_t(\boldsymbol{\Omega})\, \hat{\bx}_{t-1|t-1}, 
    \\
    \label{eq:bknet_update}
    \hat{\bx}_{t|t} &= \hat{\bx}_{t|t-1} + \bK_t(\boldsymbol{\Theta}) \big(\by_t - \bh(\hat{\bx}_{t|t-1})\big). 
\end{align}
The overall procedure of Blind-KalmanNet is summarized in Algorithm~\ref{alg:bknet}.
The key difference from KalmanNet in \eqref{eq:x_pos} is that the unavailable state-evolution function $\bff(\cdot)$ is replaced by the learned matrix $\bF_t(\boldsymbol{\Omega})$.

In each time instance $t \in \CMcal{T}$, Blind-KalmanNet estimates $\hat{\bx}_{t|t}$ via the following two steps:
\begin{enumerate}
    \item \textit{Prediction:} F-Net produces $\bF_t(\boldsymbol{\Omega})$ from history-dependent features, and the prior state is predicted as \eqref{eq:bknet_predict}. 
    The predicted observation $\hat{\by}_{t|t-1}=\bh(\hat{\bx}_{t|t-1})$ and the innovation $\by_t-\hat{\by}_{t|t-1}$ are then formed for correction.

    \item \textit{Update:} K-Net outputs $\bK_t(\boldsymbol{\Theta})$ from innovation- and state-difference features, and the posterior is updated by \eqref{eq:bknet_update}.
    This update does not require the noise covariances $\bQ$ and $\bR$ to compute the Kalman gain.
\end{enumerate}

Following the model-based deep learning paradigm \cite{shlezinger2023model}, the Blind-KalmanNet architecture retains the Kalman recursion as a principled backbone and uses DNNs only for the unavailable components $\bF_t(\boldsymbol{\Omega})$ and $\bK_t(\boldsymbol{\Theta})$.
Unlike an end-to-end network that must learn the entire filtering map from data, this model-based design confines learning to a much smaller hypothesis space, making the two heads easier to train and less prone to overfitting under end-to-end training.
Although Regression-based KalmanNet and Blind-KalmanNet use the same labeled GTD, they exploit it differently.
Regression-based KalmanNet first identifies a fixed surrogate $\widehat{\bF}$ using a one-step prediction objective, whereas Blind-KalmanNet jointly learns the state-dependent surrogate $\bF_t$ and the Kalman gain $\bK_t$ directly from the posterior state-estimation objective.

\paragraph{Neural Network Architecture}
Both heads follow the GRU-based KalmanNet architecture \#1 in \cite{revach2022kalmannet}, leveraging recurrent memory to aggregate memory-dependent information that is essential for sequential filtering.
Each head first embeds its input feature vector through a fully connected (FC) layer, and feeds the resulting representation to a multi-layer GRU.
The GRU hidden dimension is chosen proportional to $m^2+n^2$, reflecting the scale of the surrogate and statistical moments matrices implicitly tracked by the filter.
For the dimensionality of the GRU hidden states, we follow the numerical study in \cite{revach2022kalmannet}, which is $10\, (m^2 + n^2)$.
Finally, an output FC layer maps the GRU hidden state to the desired estimate: $\bF_t(\boldsymbol{\Omega}) \in\mathbb{R}^{m\times m}$ for F-Net or $\bK_t(\boldsymbol{\Theta})\in\mathbb{R}^{m\times n}$ for K-Net.

Let $\CMcal{G}_\bF(\cdot \,; \boldsymbol{\Omega})$ and $\CMcal{G}_\bK(\cdot \,; \boldsymbol{\Theta})$ denote F-Net and K-Net with their trainable parameters, respectively.
Using the DNNs, the input-output relationships are expressed as
\begin{align}
    \mathrm{vec}(\bF_t) &= \CMcal{G}_\bF\big(\boldsymbol{\delta}^{\bF}_t; \boldsymbol{\Omega}\big) \in \mathbb{R}^{m^2}, \\
    \mathrm{vec}(\bK_t) &= \CMcal{G}_\bK\big(\boldsymbol{\delta}^{\bK}_t; \boldsymbol{\Theta}\big) \in \mathbb{R}^{mn},
\end{align}
where $\boldsymbol{\delta}^{\bF}_t$ and $\boldsymbol{\delta}^{\bK}_t$ are the input feature vectors of F-Net and K-Net, respectively.

\begin{algorithm}[t]
    \DontPrintSemicolon
    \caption{Blind-KalmanNet at Time $t$}
    \label{alg:bknet}
    \KwIn{
    Observation $\mathbf{y}_t$, previous posterior estimate
    $\hat{\mathbf{x}}_{t-1|t-1}$, filtering history required to construct
    $\boldsymbol{\delta}^{\bF}_t$ and $\boldsymbol{\delta}^{\bK}_t$,
    trained F-Net$\big(\boldsymbol{\Omega}\big)$ and
    K-Net$\big(\boldsymbol{\Theta}\big)$.
    }
    
    $\mathbf{F}_t \leftarrow
    \CMcal{G}_\bF\big(\boldsymbol{\delta}^{\bF}_t;
    \boldsymbol{\Omega}\big)$
    \;

    $\hat{\mathbf{x}}_{t|t-1}
    \leftarrow
    \mathbf{F}_t\,\hat{\mathbf{x}}_{t-1|t-1}$
    and
    $\hat{\mathbf{y}}_{t|t-1}
    \leftarrow
    \mathbf{h}(\hat{\mathbf{x}}_{t|t-1})$\;

    $\Delta\mathbf{y}_t
    \leftarrow
    \mathbf{y}_t-\hat{\mathbf{y}}_{t|t-1}$\;

    $\hat{\mathbf{x}}_{t|t}
    \leftarrow
    \hat{\mathbf{x}}_{t|t-1}
    +
    \mathbf{K}_t\Delta\mathbf{y}_t$
    where
    $\mathbf{K}_t
    \leftarrow
    \CMcal{G}_\bK\big(
    \boldsymbol{\delta}^{\bK}_t;
    \boldsymbol{\Theta}\big)$\;

\end{algorithm}

\subsection{Input Features}
We construct scale-invariant features via $\ell_2$ normalization.
For any vector $\ba$, let $\cN(\ba) \triangleq \ba/(\|\ba\|+\epsilon)$ with a small $\epsilon>0$; each feature defined below is normalized by $\cN(\cdot)$ before being fed into the networks.

\textbf{K-Net features.}
Following  KalmanNet~\cite{revach2022kalmannet}, the K-Net input is the concatenation of four difference-based features:
\begin{align}
    \label{eq:KNet_features}\boldsymbol{\delta}^{\bK}_t =
    \Big[
        \Delta\tilde{\by}_t^{\sf T},\;
        \Delta\by_t^{\sf T},\;
        \Delta\tilde{\bx}_{t-1}^{\sf T},\;
        \Delta\hat{\bx}_t^{\sf T}
    \Big]^{\sf T},
\end{align}
where $\Delta\tilde{\by}_t = \by_t - \by_{t-1}$ is the observation difference,
$\Delta\by_t = \by_t - \hat{\by}_{t|t-1}$ is the innovation,
$\Delta\tilde{\bx}_{t-1} = \hat{\bx}_{t-1|t-1} - \hat{\bx}_{t-2|t-2}$
is the forward evolution difference,
and $\Delta\hat{\bx}_t = \hat{\bx}_{t|t-1} - \hat{\bx}_{t-1|t-1}$
is the forward update difference.
These features jointly summarize the local innovation and the recent state-update behavior.

\textbf{F-Net features.}
Unlike K-Net, F-Net produces a surrogate $\bF_t(\boldsymbol{\Omega})$ that describes how the state evolves, and thus requires features that identify the current
operating point rather than differences alone.
Accordingly, we feed F-Net the recent observations and posterior estimates:
\begin{align}
    \label{eq:FNet_input}
    \boldsymbol{\delta}^{\bF}_t =
    \Big[
        \by_t^{\sf T},\;
        \by_{t-1}^{\sf T},\;
        \hat{\bx}_{t-1|t-1}^{\sf T},\;
        \hat{\bx}_{t-2|t-2}^{\sf T}
    \Big]^{\!\sf T}.
\end{align}
The pairs $(\by_t,\by_{t-1})$ and
$(\hat{\bx}_{t-1|t-1},\hat{\bx}_{t-2|t-2})$ provide F-Net with information about both the current state of the trajectory and its recent evolution, allowing it to infer a local state transition.
Using the absolute observations and state estimates, rather than only their differences, enables the learned surrogate to depend on the current location
along the trajectory.
Temporal changes in the surrogate are further captured by the RNN memory.

\subsection{Training Blind-KalmanNet}
The proposed Blind-KalmanNet is trained in a supervised, end-to-end manner using labeled trajectories. Unlike KalmanNet, which only learns the Kalman gain $\bK_t(\boldsymbol{\Theta})$, Blind-KalmanNet simultaneously needs to learn both  $\bF_t(\boldsymbol{\Omega})$ and  $\bK_t(\boldsymbol{\Theta})$.
Since the posterior state estimate $\hat{\bx}_{t|t}$ takes values in the continuous space $\bbR^{m}$, we adopt the squared-error loss:
\begin{align}
    \label{eq:bknet_loss}
    \CMcal{L} = \|\bx_t - \hat{\bx}_{t|t}\|^2.
\end{align}
Note that although the internal RNN modules output $\bF_t(\boldsymbol{\Omega})$ and $\bK_t(\boldsymbol{\Theta})$ (rather than $\hat{\bx}_{t|t}$ directly), the loss is computed on the final state estimate $\hat{\bx}_{t|t}$, enabling gradient-based learning through the entire filtering computation graph.

In \cite{revach2022kalmannet}, it was shown that the Kalman gain is trainable via backpropagation through the MSE loss.
We extend this result to verify the trainability of the state transition matrix $\bF_t$ in Blind-KalmanNet.
Recall from \eqref{eq:bknet_predict} that the priori estimate is given by $\hat{\bx}_{t|t-1} = \bF_t\, \hat{\bx}_{t-1|t-1}$.
For clarity, we derive the partial derivative $\partial \CMcal{L}/\partial \bF_t$ while holding $\bK_t$ fixed, which isolates the direct effect of $\bF_t$ on the prediction step:
\begin{align}
    \nonumber
    \frac{\partial \CMcal{L}}{\partial\bF_t} &= \frac{\partial \CMcal{L}}{\partial \hat{\bx}_{t|t-1}} \, \hat{\bx}_{t-1|t-1}^{\sf T} 
    \\ \label{eq:gd_BKNet}
    &= 2\left(\bI - \bK_t {\bH}_t \right)^{\sf T}  \left(\hat{\bx}_{t|t}  - \bx_t \right)\hat{\bx}_{t-1|t-1}^{\sf T},
\end{align}
where ${\bH}_t$ is the Jacobian matrix of the nonlinear observation function.
The gradient in \eqref{eq:gd_BKNet} confirms that $\bF_t$ can be learned through end-to-end training, and it is well-defined and non-trivial whenever $\hat{\bx}_{t-1|t-1} \neq {\bf 0}$.
In end-to-end training, however, gradients are backpropagated through the full computation graph, and any indirect dependence of $\bK_t(\boldsymbol{\Theta})$ on $\bF_t(\boldsymbol{\Omega})$ through the K-Net inputs is automatically accounted for.

We train both F-Net and K-Net jointly by minimizing the MSE loss over the training trajectories. 
Let $\boldsymbol{\CMcal{D}}$ denote the index set of training sequences of length $T$.
The optimization problem is formulated as:
\begin{align}
    \left(\boldsymbol{\Omega}^{\star}\!, \boldsymbol{\Theta}^{\star}\right) \!= \argmin_{\boldsymbol{\Omega}, \boldsymbol{\Theta}} \frac{1}{T|\boldsymbol{\CMcal{D}}|} \sum_{\ell \in \boldsymbol{\CMcal{D}}} \sum_{t=1}^{T} \left\|\bx_t^{(\ell)} - \hat{\bx}_{t|t}^{(\ell)} \right\|^2\!\!.
\end{align}
At each time step $t$, F-Net first computes the state transition matrix $\bF_t(\boldsymbol{\Omega})$  from the input features $\boldsymbol{\delta}_t^{\bF}$, which is then used to compute the prior estimate $\hat{\bx}_{t|t-1}$.
Subsequently, K-Net computes the Kalman gain $\bK_t(\boldsymbol{\Theta})$ based on the innovation-related features $\boldsymbol{\delta}_t^{\bK}$.
This sequential computation naturally couples the two networks, as the prior estimate produced by F-Net directly influences the innovation used by K-Net.

Gradients are backpropagated through the entire computation graph using backpropagation through time (BPTT) as in \cite{revach2022kalmannet}, including the coupling between F-Net outputs and K-Net inputs.
This joint training allows both networks to co-adapt:  F-Net learns to produce prior estimates that, when combined with the learned  Kalman gain from K-Net, minimize the overall sequential state estimation error. 
In practice, we initialize the GRU hidden states to zeros at the beginning of each trajectory and process each trajectory sequentially through time.

Blind-KalmanNet realizes the task-aware principle in its most direct form.
Both the predictor and the corrector are learned from the posterior state-estimation objective.
Its predictor, however, remains a state-dependent linear surrogate.
Accordingly, unknown dynamics (possibly nonlinear) is represented through a sequence of local linear approximations rather than an explicit globally structured
dynamics model.
To complement this local representation, Koopman operator theory provides a lifted space in which nonlinear models can be represented approximately by a globally shared linear operator.
This motivates the second stage of the design.
The next section develops Blind-Koopman-KalmanNet, which replaces the purely local surrogate with a pre-trained Koopman predictor serving as a globally structured backbone, while retaining the task-aware learning of Blind-KalmanNet in the form of a residual surrogate and a learned Kalman gain.

\section{Blind-KalmanNet with Koopman Lifting}\label{sec:BK2Net}

We now present Koopman-aided Blind-KalmanNet, referred to as \emph{Blind-Koopman-KalmanNet}, which completes the design initiated in Section~\ref{sec:BKNet}.
Blind-KalmanNet retains the task-aware principle but represents the unknown dynamics through a purely local, unstructured predictor.
To complement this local representation, we introduce Koopman operator theory, which provides a lifted space in which nonlinear dynamics can be approximated by a globally shared linear operator.
This lifted representation provides a structured backbone for prediction while retaining the task-aware adaptation developed in Blind-KalmanNet.
This added structure, however, comes at the cost of a two-stage training pipeline and a larger model, and the resulting trade-off between the two proposed frameworks is examined in Section~\ref{sec:simulation}.

\subsection{Koopman Operator} \label{subsec:Koopman}
Koopman operator theory provides a principled way to represent nonlinear dynamics through a linear (generally infinite-dimensional) operator acting on a space of observables.
Let $g:\mathcal{X}\rightarrow\mathbb{R}$ be an observable in an infinite-dimensional Hilbert space $\bbH$.
The Koopman operator $\cK: \bbH \rightarrow \bbH$ advances observables linearly via
\begin{align}
   \cK g(\bx_t) = g\big(\bff(\bx_t)\big) = g(\bx_{t+1}).
\end{align}
The infinite-dimensionality of $\bbH$ poses challenges in practical Koopman analysis, motivating a finite-dimensional approximation within an invariant subspace.

A natural choice is to span such an invariant subspace by the eigenfunctions $\{\varphi_1,\varphi_2,\ldots,\varphi_M\}$ of $\cK$ \cite{brunton2016koopman}, under which the action of $\cK$ remains closed.
The resulting linearized dynamics is expressed as
\begin{align}
    \boldsymbol{\varphi}(\bx_{t+1})
    =
    \boldsymbol{\CMcal{K}}\boldsymbol{\varphi}(\bx_t),
\end{align}
where $\boldsymbol{\CMcal{K}}\in\bbR^{M\times M}$ is the finite-dimensional Koopman matrix and $\boldsymbol{\varphi}$ is the $M$-dimensional mapping whose elements are the eigenfunctions of $\cK$.
Consequently, the state evolution in the original space is approximated as
\begin{align}
    \bx_{t+1}
    \approx
    \boldsymbol{\varphi}^{-1}\!\big(
    \boldsymbol{\CMcal{K}}\boldsymbol{\varphi}(\bx_t)
    \big),
\end{align}
where $\boldsymbol{\varphi}^{-1}$ denotes the inverse mapping (decoding) back to the original space.

In our scenarios, applying the Koopman framework to KalmanNet is conceptually straightforward as a way to mimic $\bff(\cdot)$.
However, identifying $\boldsymbol{\CMcal{K}}$ and $(\boldsymbol{\varphi},\boldsymbol{\varphi}^{-1})$ for a general nonlinear system is challenging, even for data-driven approaches \cite{lusch2018deep}.
We therefore adopt a deep learning-based approach to learn these components, as detailed in the next subsection.

\subsection{Blind-Koopman-KalmanNet Architecture}

To leverage Koopman operator theory, we adopt a DKN~
\cite{takeishi2017learning,lusch2018deep,shi2022deep} using the available GTD.
The DKN learns $(i)$ an encoder $\phi$ that lifts the state into a latent Koopman space,
$(ii)$ a Koopman matrix $\boldsymbol{\CMcal{K}}$ that propagates the lifted state linearly,
and $(iii)$ a decoder $\psi$ that maps the lifted state back to the original space,
following the architectural choices in~\cite{chen2025kalman}.
For a chosen latent dimension $D_z\geq m$, we have
\begin{align}
\bz_t &= \phi(\bx_t)\in\bbR^{D_z},
&
\bz_{t+1} &= \boldsymbol{\CMcal{K}}\bz_t,
&
\bx_{t+1} &\approx \psi(\bz_{t+1}).
\end{align}

Here, $\boldsymbol{\CMcal{K}}\in\bbR^{D_z\times D_z}$ provides a globally shared
linear representation of the dynamics in the lifted space.
Since the filtering recursion is performed in the original state space, we
adopt a state-augmented encoder that preserves direct access to the original state:
\begin{align}
\label{eq:state_aug}
\phi(\bx)
\triangleq
\begin{bmatrix}
\bx\\
\bg(\bx)
\end{bmatrix}
\in\bbR^{D_z},
\end{align}
where $\bg:\bbR^m\rightarrow\bbR^{D_z-m}$ is a feedforward DNN.

We define the state-selection matrix
\begin{align}
\label{eq:state_projection}
\bP_m
\triangleq
\begin{bmatrix}
\bI_m & \boldsymbol{0}
\end{bmatrix}
\in\bbR^{m\times D_z}.
\end{align}
Accordingly, the decoder is given by the fixed projection
\begin{align}
\label{eq:decoder}
\psi(\bz)
\triangleq
\bP_m\bz
=
\bz_{1:m},
\end{align}
which satisfies
$\psi(\phi(\bx))=\bP_m\phi(\bx)=\bx$.
We suppress the parameters of $\phi$ and $\boldsymbol{\CMcal{K}}$ in the filtering
notation.
These components are pre-trained and remain fixed during filtering.

Since the first $m$ latent coordinates correspond directly to the original state,
we define the original-space Koopman predictor:
\begin{align}
\label{eq:K_m}
\boldsymbol{\CMcal{K}}_m
\triangleq
\bP_m\boldsymbol{\CMcal{K}}
=
\big[\boldsymbol{\CMcal{K}}\big]_{1:m}
\in\bbR^{m\times D_z}.
\end{align}
The full $D_z\times D_z$ Koopman matrix is learned to promote a globally linear
latent evolution, whereas $\boldsymbol{\CMcal{K}}_m$ maps the lifted state to
the corresponding one-step prediction in the original state space.

\begin{figure}[t]
    {\centerline{
    \resizebox{1\columnwidth}{!}{
    \includegraphics{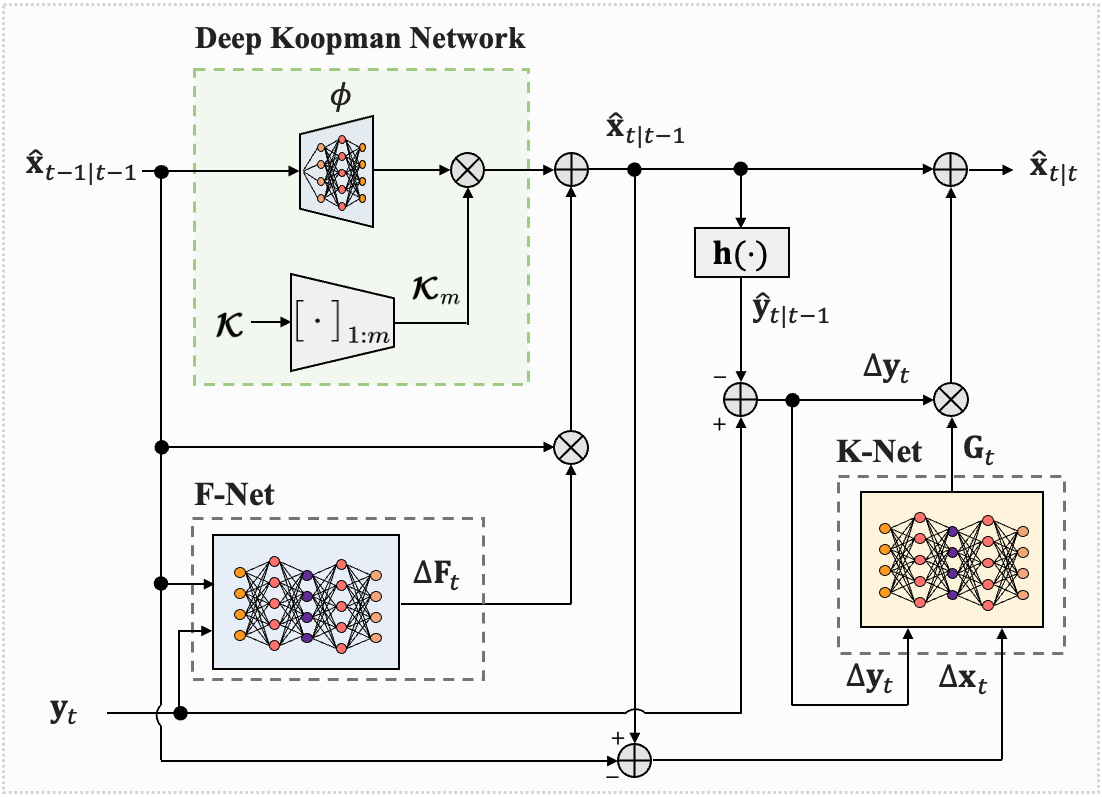}
    }}
    \caption{Blind-Koopman-KalmanNet block diagram.}
    \label{fig:B_K2Net}}
\end{figure}

The mismatch between one-step DKN pre-training and the downstream filtering
objective, together with the approximate nature of a finite-dimensional Koopman
representation, may result in residual prediction errors during filtering.
We therefore augment the Koopman predictor with a task-aware residual surrogate
learned directly from the filtering objective.
Specifically, an additional residual F-Net produces
\begin{align}
\label{eq:delta_F_K2Net}
\operatorname{vec}\left(
\Delta\bF_t\left(\bar{\bOmega}\right)
\right)
=
\CMcal{G}_{\bF}
\left(
\bar{\boldsymbol{\delta}}^{\bF}_t;
\bar{\bOmega}
\right)
\in\bbR^{m^2},
\end{align}
where $\bar{\bOmega}$ denotes the residual F-Net parameters, $\bar{\boldsymbol{\delta}}^{\bF}_t$ denotes the input feature vector defined in \eqref{eq:FNet_input}, and
$\Delta\bF_t(\bar{\bOmega})\in\bbR^{m\times m}$.
The residual F-Net follows the same GRU-based architecture as in Blind-KalmanNet.
This completes the Blind-Koopman-KalmanNet architecture, whose overall structure is depicted in Fig.~\ref{fig:B_K2Net}.
We next describe its main components in detail.

\paragraph{Koopman Encoder $\phi$}
The state-augmented encoder in \eqref{eq:state_aug} preserves the original state
in the first $m$ latent coordinates and augments it with nonlinear observables
$\bg(\bx)$.
This structure provides direct access to the original state while allowing the
additional latent coordinates to capture features useful for representing the
underlying dynamics.

\paragraph{Koopman Matrix $\boldsymbol{\CMcal{K}}$}
The matrix $\boldsymbol{\CMcal{K}}$ approximates the Koopman operator and
propagates the lifted representation linearly, while its top-$m$ rows
$\boldsymbol{\CMcal{K}}_m$ map the lifted state to a one-step prediction in the
original state space.
The same Koopman matrix is shared across all time steps and trajectories and
remains fixed during filtering, thereby providing a globally structured
prediction backbone.
Filtering-specific adaptation is instead handled by the residual F-Net.

\paragraph{Residual F-Net}
The residual F-Net generates $\Delta\bF_t(\bar{\bOmega})$ from the filtering
history.
Its role is to adapt the pre-trained Koopman predictor to the filtering task in a task-aware manner.
Accordingly, $\boldsymbol{\CMcal{K}}_m\phi(\cdot)$ provides the globally
structured prediction, while
$\Delta\bF_t(\bar{\bOmega})\hat{\bx}_{t-1|t-1}$ provides state-dependent,
task-aware local adaptation inherited from Blind-KalmanNet.

\paragraph{Koopman Decoder $\psi$}
Under the state-augmented encoder, the decoder reduces to the fixed projection
$\psi(\bz)=\bz_{1:m}$ and therefore requires no separately trained network.
The posterior is maintained in the original state space and re-lifted through
$\phi$ at each time step, rather than propagating a separate latent posterior
across time.

\vspace{0.6em}

We now describe the enhanced prediction step with the task-aware residual F-Net.

\textbf{Prediction.}
Given the previous posterior $\hat{\bx}_{t-1|t-1}$, the Koopman branch and the residual F-Net jointly form the prior:
\begin{align}
    \label{eq:x_prior_K2Net}
    \hat{\bx}_{t|t-1}
    &=
    \boldsymbol{\CMcal{K}}_m
    \phi\big(\hat{\bx}_{t-1|t-1}\big)
    +
    \Delta\bF_t(\bar{\bOmega})
    \hat{\bx}_{t-1|t-1},
    \\
    \label{eq:y_prior_K2Net}
    \hat{\by}_{t|t-1}
    &=
    \bh\big(\hat{\bx}_{t|t-1}\big),
    \\
    \label{eq:innovation_K2Net}
    \Delta\by_t
    &=
    \by_t-\hat{\by}_{t|t-1}.
\end{align}
Using $\bP_m\phi(\bx)=\bx$, the predictor in \eqref{eq:x_prior_K2Net} can equivalently be written as
\begin{align}
    \nonumber
    \hat{\bx}_{t|t-1}
    =
    \left(
    \boldsymbol{\CMcal{K}}_m
    +
    \Delta\bF_t(\bar{\bOmega})\bP_m
    \right)
    \phi\big(\hat{\bx}_{t-1|t-1}\big) =  \boldsymbol{\CMcal{K}}_{m,t}^{\rm eff} \phi\big(\hat{\bx}_{t-1|t-1}\big),
\end{align}
where $\boldsymbol{\CMcal{K}}_{m,t}^{\rm eff}  = \boldsymbol{\CMcal{K}}_m
    +    \Delta\bF_t(\bar{\bOmega})\bP_m$. The variation of $\boldsymbol{\CMcal{K}}_{m,t}^{\rm eff}$ across filtering steps
arises solely from the task-aware residual, whereas the pre-trained Koopman
predictor $\boldsymbol{\CMcal{K}}_m$ remains globally shared and fixed.
The residual F-Net recomputes $\Delta\bF_t(\bar{\bOmega})$ from the recent
filtering history  and is optimized through the
posterior state-estimation objective.

\textbf{Update (K-Net).}
The K-Net outputs a gain $\bG_t(\bar{\bTheta})\in\bbR^{m\times n}$ and updates the posterior according to
\begin{align}
    \label{eq:x_post_K2Net}
    \hat{\bx}_{t|t}
    =
    \hat{\bx}_{t|t-1}
    +
    \bG_t(\bar{\bTheta})\Delta\by_t.
\end{align}
The K-Net uses $\bar{\boldsymbol{\delta}}^{\bK}_t$ that contains the same input features as in \eqref{eq:KNet_features}, evaluated using the residual-corrected prior in  \eqref{eq:x_prior_K2Net}.
The residual F-Net and K-Net are jointly trained, while the DKN backbone remains fixed.
The overall procedure is summarized in Algorithm~\ref{alg:bk2net}.

\begin{algorithm}[t]
\DontPrintSemicolon
\caption{Blind-Koopman-KalmanNet at Time $t$}
\label{alg:bk2net}
\KwIn{
Observation $\mathbf{y}_t$, previous posterior estimate $\hat{\mathbf{x}}_{t\!-\!1|t\!-\!1}$, filtering history required to construct $\bar{\boldsymbol{\delta}}^{\mathbf{F}}_t$ and $\bar{\boldsymbol{\delta}}^{\mathbf{K}}_t$, pre-trained and fixed DKN $(\phi,\boldsymbol{\CMcal{K}}_m)$, trained F-Net $(\bar{\bOmega})$, and trained K-Net $(\bar{\bTheta})$.
}

$\Delta\mathbf{F}_t(\bar{\bOmega})
\leftarrow
\CMcal{G}_{\mathbf{F}}
\big(
\bar{\boldsymbol{\delta}}^{\mathbf{F}}_t;
\bar{\bOmega}
\big)$\;

$\hat{\mathbf{x}}_{t|t-1}
\leftarrow
\boldsymbol{\CMcal{K}}_m
\phi(\hat{\mathbf{x}}_{t-1|t-1})
+
\Delta\mathbf{F}_t(\bar{\bOmega})
\hat{\mathbf{x}}_{t-1|t-1}$\;

$\hat{\mathbf{y}}_{t|t-1}
\leftarrow
\mathbf{h}(\hat{\mathbf{x}}_{t|t-1})$\;

$\Delta\mathbf{y}_t
\leftarrow
\mathbf{y}_t-\hat{\mathbf{y}}_{t|t-1}$\;

$\mathbf{G}_t(\bar{\bTheta})
\leftarrow
\CMcal{G}_{\mathbf{K}}
\big(
\bar{\boldsymbol{\delta}}^{\mathbf{K}}_t;
\bar{\bTheta}
\big)
\in\bbR^{m\times n}$\;

$\hat{\mathbf{x}}_{t|t}
\leftarrow
\hat{\mathbf{x}}_{t|t-1}
+
\mathbf{G}_t(\bar{\bTheta})
\Delta\mathbf{y}_t$\;

\end{algorithm}

\subsection{Training Blind-Koopman-KalmanNet}

We train Blind-Koopman-KalmanNet using a two-stage strategy:
$(i)$ the DKN is pre-trained from consecutive ground-truth state pairs to learn a task-independent Koopman predictor, and
$(ii)$ the DKN is frozen while the residual F-Net and K-Net are jointly trained on labeled state-observation trajectories using the posterior state-estimation objective.
Freezing the DKN preserves the pre-trained Koopman representation, while filtering-specific adaptation is delegated to the residual F-Net.

\vspace{0.5em}

\textbf{Stage 1: Task-independent DKN pre-training.}
From the GTD trajectories
$\{\bx^{(\ell)}_{1:T}\},\;\ell\in\boldsymbol{\CMcal{D}}$,
we form one-step pairs $(\bx_t,\bx_{t+1})$ and learn the DKN components
$(\phi,\boldsymbol{\CMcal{K}})$ by minimizing a weighted sum of the lifted linear-dynamics loss and the original-space prediction loss:
\begin{align}
    \cL_{\rm DKN}^{\rm d}
    &=
    \big\|
    \phi(\bx_{t+1})
    -
    \boldsymbol{\CMcal{K}}
    \phi(\bx_t)
    \big\|^2,
    \\
    \cL_{\rm DKN}^{\rm p}
    &=
    \big\|
    \bx_{t+1}
    -
    \boldsymbol{\CMcal{K}}_m
    \phi(\bx_t)
    \big\|^2.
\end{align}

The dynamics loss $\cL_{\rm DKN}^{\rm d}$ encourages an approximately linear evolution of the full lifted state, while the prediction loss $\cL_{\rm DKN}^{\rm p}$ directly supervises the original-space Koopman branch used during filtering.
Under the state-augmented encoder and fixed projection decoder, $\bx_t-\psi(\phi(\bx_t))=\boldsymbol{0}$ by construction; hence, no separate
reconstruction loss is required.
Using weights $(w_{\rm d},w_{\rm p})$ to balance the two objectives, the combined DKN loss is defined as $\cL_{\rm DKN}
    =
    w_{\rm d}\cL_{\rm DKN}^{\rm d}
    +
    w_{\rm p}\cL_{\rm DKN}^{\rm p}$.
Accordingly, the pre-trained DKN is obtained as
\begin{align}
    \label{eq:DKN_pretraining_opt}
    \big(
    \phi^\star,
    \boldsymbol{\CMcal{K}}^\star
    \big)
    =
    \argmin_{\phi,\boldsymbol{\CMcal{K}}}
    \cL_{\rm DKN}.
\end{align}
After Stage~1, $\phi^\star$ and $\boldsymbol{\CMcal{K}}^\star$ are frozen and excluded from the Stage~2 optimizer.

Although we do not explicitly constrain the spectral radius of
$\boldsymbol{\CMcal{K}}$, the proposed filter does not perform a multi-step
latent rollout by repeatedly applying the Koopman matrix.
At each time step, the posterior estimate is corrected using the current
observation, re-lifted into the latent space, and
$\boldsymbol{\CMcal{K}}_m$ is applied only once for the next prediction.
Therefore, the long-term behavior of the estimator is governed by the overall
predictor--corrector recursion rather than by repeated powers of
$\boldsymbol{\CMcal{K}}$ alone.

\vspace{0.5em}

\textbf{Stage 2: Task-aware residual F-Net and K-Net learning.}
With the pre-trained DKN fixed, we jointly train the residual F-Net parameters $\bar{\bOmega}$ and K-Net parameters $\bar{\bTheta}$.
For each training trajectory $\ell\in\boldsymbol{\CMcal{D}}$, the filtering recursion is unrolled as
\begin{align}
    \Delta\bF_t^{(\ell)}(\bar{\bOmega})
    &=
    \CMcal{G}_{\bF}
    \big(
    \bar{\boldsymbol{\delta}}_{t}^{\bF,(\ell)};
    \bar{\bOmega}
    \big),
    \\
    \hat{\bx}_{t|t-1}^{(\ell)}
    &=
    \boldsymbol{\CMcal{K}}_m^\star\,
    \phi^\star\big(
    \hat{\bx}_{t-1|t-1}^{(\ell)}
    \big)
    +
    \Delta\bF_t^{(\ell)}(\bar{\bOmega})\,
    \hat{\bx}_{t-1|t-1}^{(\ell)},
    \\
    \hat{\by}_{t|t-1}^{(\ell)}
    &=
    \bh\big(
    \hat{\bx}_{t|t-1}^{(\ell)}
    \big),
    \\
    \hat{\bx}_{t|t}^{(\ell)}
    &=
    \hat{\bx}_{t|t-1}^{(\ell)}
    +
    \bG_t(\bar{\bTheta})
    \big(
    \by_t^{(\ell)}
    -
    \hat{\by}_{t|t-1}^{(\ell)}
    \big),
\end{align}
initialized with $\hat{\bx}_{0|0}^{(\ell)}$.

The residual F-Net and K-Net are jointly optimized using the trajectory-averaged
empirical MSE:
\begin{align}
    \label{eq:loss_task_K2Net}
    \cL_{\rm task}\big(
    \bar{\bOmega},
    \bar{\bTheta};
    \phi^\star,
    \boldsymbol{\CMcal{K}}^\star
    \big)
    =
    \frac{1}{
    {T}
    |\boldsymbol{\CMcal{D}}|
    }
    \sum_{\ell\in\boldsymbol{\CMcal{D}}}
    \sum_{t=1}^{T}
    \left\|
    \bx_t^{(\ell)}
    -
    \hat{\bx}_{t|t}^{(\ell)}
    \right\|^2.
\end{align}

Consequently, the Stage~2 optimization is 
\begin{align}
    \label{eq:loss_stage2_K2Net}
    \big(
    \bar{\bOmega}^\star,
    \bar{\bTheta}^\star
    \big)
    =
    \argmin_{\bar{\bOmega},\bar{\bTheta}}
    \cL_{\rm task}
    \big(
    \bar{\bOmega},
    \bar{\bTheta};
    \phi^\star,
    \boldsymbol{\CMcal{K}}^\star
    \big).
\end{align}
Gradients are backpropagated through the unrolled filtering recursion via BPTT to the residual F-Net and K-Net only; the frozen DKN backbone receives no gradient update.

\section{Simulation Results} \label{sec:simulation}
In this section, we present an extensive numerical study of the proposed Blind-KalmanNet frameworks, in the unknown-dynamics setting.
We evaluate the state-estimation performance under linear and nonlinear SS models as well as real-world dynamics, and compare them with both model-based and data-driven baselines.

\subsection{Experimental Setting}
The proposed methods operate without knowledge of $\bff(\cdot)$, $\bQ$, and
$\bR$. 
The competing baselines are provided with the additional model
information explicitly specified below:
\begin{itemize}
    \item \textbf{B-KNet}: Proposed Blind-KalmanNet in Section~\ref{sec:BKNet}.
    
    \item \textbf{B-K$^{\mathbf{2}}$Net}: Proposed Blind-Koopman-KalmanNet  in Section~\ref{sec:BK2Net}.
    
    \item \textbf{Reg-KNet}: Regression-based KalmanNet in Section~\ref{subsec:Reg_KNet}.
    
    \item \textbf{Reg-EKF}: Regression-based EKF that uses the same regressed matrix $\widehat{\bF}$ as Reg-KNet, while assuming $\bQ$ and $\bR$ are known.
    
    \item \textbf{BKF}: BKF in \cite{sharma2020blind}, while assuming $\bQ$ and $\bR$ are known.
    
    \item \textbf{AKF}: AKF with $\bff(\cdot)=\bI$, where $\bQ$ is estimated online \cite{myers1976adaptive} and $\bR$ is known.

    \item \textbf{RNN}: End-to-end RNN trained directly to map $\by_{t}$ to $\hat{\bx}_t$, without any Kalman structure.

    \item \textbf{Mamba}: A selective SS model-based deep learning approach  \cite{gu2023mamba}, which is trained to directly map the measurement history $\by_{1:t}$ to the state estimate $\hat{\bx}_t$.

    \item \textbf{DANSE}: The original DANSE~\cite{ghosh2024danse} for linear observations and pDANSE~\cite{ghosh2025pdanse} with $50$ particles for nonlinear ones. 
    Both require $\bR$ and are trained with full supervision on the same GTD as all other AI-aided schemes.

    \item \textbf{EKF (perfect)}: EKF that has full knowledge of $\bff(\cdot)$, $\bQ$, and $\bR$, serving as an oracle model-based baseline.

    \item \textbf{KNet (perfect)}: KalmanNet with full knowledge of $\bff(\cdot)$, serving as an oracle AI-aided baseline.
\end{itemize}
Here, we note that all AI-aided schemes are trained on the same GTD.
For linear SS models, the EKF-based schemes reduce to the standard KF, since the Jacobian matrices coincide with the linear dynamics $\bF$ or $\bH$.

For each experiment, the dataset is split into disjoint sets of $N_E = 100$ training, $N_V = 10$ validation, and $N_T = 10$ test trajectories, unless mentioned otherwise.
Model parameters are optimized using mini-batches of size $10$.
Training and validation use trajectories of length $T_{\rm train}$, while test sequences have length $T_{\rm test}$.
We evaluate the state-estimation accuracy by empirical MSE averaged over test trajectories:
\begin{align}
    \mathrm{MSE} = \frac{1}{|\boldsymbol{\CMcal{D}}_{\rm test}|\, T_{\rm test}} \sum_{\ell \in \boldsymbol{\CMcal{D}}_{\rm test}} \sum_{t=1}^{T_{\rm test}} \big\| \bx_t^{(\ell)} - \hat{\bx}_{t|t}^{(\ell)} \big\|^2,
    \label{eq:MSE}
\end{align}
where $\boldsymbol{\CMcal{D}}_{\rm test}$ denotes the index set of test trajectories. 

The state vector $\bx_t$ is represented in Cartesian coordinates, e.g., $\bx_t = [x_{1,t}, x_{2,t}, x_{3,t}]^{\sf T}$ for $m=3$.
Unless stated otherwise, the SS model is generated with isotropic process and observation noise covariances:
\begin{equation}
    \label{eq:noise_cov}
    \bQ = \sigma_{\rm q}^2 \bI, \quad \bR = \sigma_{\rm r}^2 \bI.
\end{equation}
For the synthetic experiments, we adopt Gaussian process and observation noise, following standard evaluation practice.
Although Gaussian noise is used in these controlled settings, the proposed
methods do not rely on this assumption.
We further evaluate the proposed methods on the real-world dataset in Section~\ref{subsec:realworld}, which exhibits non-Gaussian noise arising from practical sensing non-idealities.

For B-K$^2$Net, we set the encoder $\bg(\cdot)$ to be a 2-layer feedforward network of hidden width $64$ with ReLU activations, set the latent dimension to $D_z = 2m$, and use $5000$ epochs for pre-training the DKN.
The DKN loss weights are $w_{\rm d} = w_{\rm p} = 1$.
All DNNs are trained with the Adam optimizer with learning rate $10^{-3}$, weight decay $10^{-4}$ in conjunction with a learning rate scheduler.
All experiments are conducted on a workstation equipped with Intel i9-13900K CPU, RTX 4080 GPU, and 64\,GB RAM, running Python 3.11.8 and PyTorch 2.11.0 with CUDA 13.0.

\subsection{Linear Dynamics}
\begin{table}[t]
\centering
\caption{{Numerical Results for the UCM example with $T_{\rm test}=80$ and $N_T = 10$.}}
\vspace{-0.5em}
\label{tab:UCM}
\renewcommand{\arraystretch}{1.15}
\setlength{\tabcolsep}{5pt}
\begin{tabular}{lcccc}
\toprule
Method
& MSE
& Ratio
& Run time [s]
& Total params. \\
\midrule

EKF (perfect)
& $-25.120$
& $100.0\%$
& $0.0932$
& -- \\

\midrule

AKF
& $-12.518$
& $49.8\%$
& $0.9341$
& -- \\

BKF
& $\bf{-25.120}$
& $\bf{100.0\%}$
& $3.1436$
& -- \\

RNN
& $-6.880$
& $27.4\%$
& $\bf{0.0344}$
& $137{,}650$ \\

Mamba
& $-24.059$
& $95.8\%$
& $0.0693$
& $39{,}026$ \\

DANSE
& $-24.846$
& $98.9\%$
& $0.0680$
& $\mathbf{5{,}176}$ \\

Reg-EKF
& $-25.080$
& $99.8\%$
& $0.0841$
& -- \\

Reg-KNet
& $-25.017$
& $99.6\%$
& $0.0619$
& $139{,}604$ \\

\midrule

\textbf{B-KNet}
& $-24.575$
& $97.8\%$
& $0.1005$
& $279{,}208$ \\

\textbf{B-K$^{2}$Net}
& $\bf{-25.120}$
& $\bf{100.0\%}$
& $0.1095$
& $396{,}696$ \\

\bottomrule
\end{tabular}
\end{table}

We first consider linear dynamics, where the KF attains the MMSE and thus serves as an optimal benchmark under a linear SS model. 
We adopt  unit circle motion (UCM) for $m=2$:
\begin{align}
    \label{eq:UCM_x}
    \bF^{\rm UCM}  = 
    \begin{pmatrix}
    \cos \theta & -\sin \theta \\
    \sin \theta & \cos \theta
    \end{pmatrix},
\end{align}
where $\theta$ is a constant rotation angle, and a linear observation model $\bh(\cdot) = \bI$ with $n=2$. 
We set the process noise variance to $\sigma_{\rm q}^2 = 10^{-3}$, and generate trajectories of length $T_{\rm train} = 40$ and $T_{\rm test} = 80$ with $\theta = \pi/10$.

Table~\ref{tab:UCM} summarizes the MSE and run time of all schemes at $1/\sigma_{\rm r}^2 = 20$ dB.
Since the UCM example is linear and time-invariant, $\bF$ can be accurately estimated from training data via linear regression.
Hence, the methods exploiting this estimate nearly match the optimal filter.
The proposed B-KNet and B-K$^2$Net also track the optimal benchmark closely, with the small remaining gap reflecting the fact that they learn the filtering map directly from data without relying on a pre-estimated $\widehat{\bF}$.
This advantage is not exercised when a single constant $\bF$ already describes the entire dynamics.

In terms of model size, as shown in Table~\ref{tab:UCM},  B-K$^2$Net has the largest parameter count among the learned schemes.
Despite this high complexity, it matches the optimal EKF benchmark while incurring only a modest increase in run time.
DANSE is the most compact learned model, as it predicts the prior moments and computes the measurement update analytically rather than learning a separate correction network.
Its run time nevertheless remains comparable to that of Mamba, since the closed-form update requires inversion of the innovation covariance at every time step.
Overall, this example confirms that the proposed methods remain competitive with the optimal filter under linear dynamics.
However, the restrictive SS model of UCM does not fully showcase their advantage, which becomes more apparent in the more general settings considered next.

\subsection{Nonlinear Dynamics}
\begin{figure}[t]    
    {\centerline{\resizebox{0.88 \columnwidth}{!}{\includegraphics{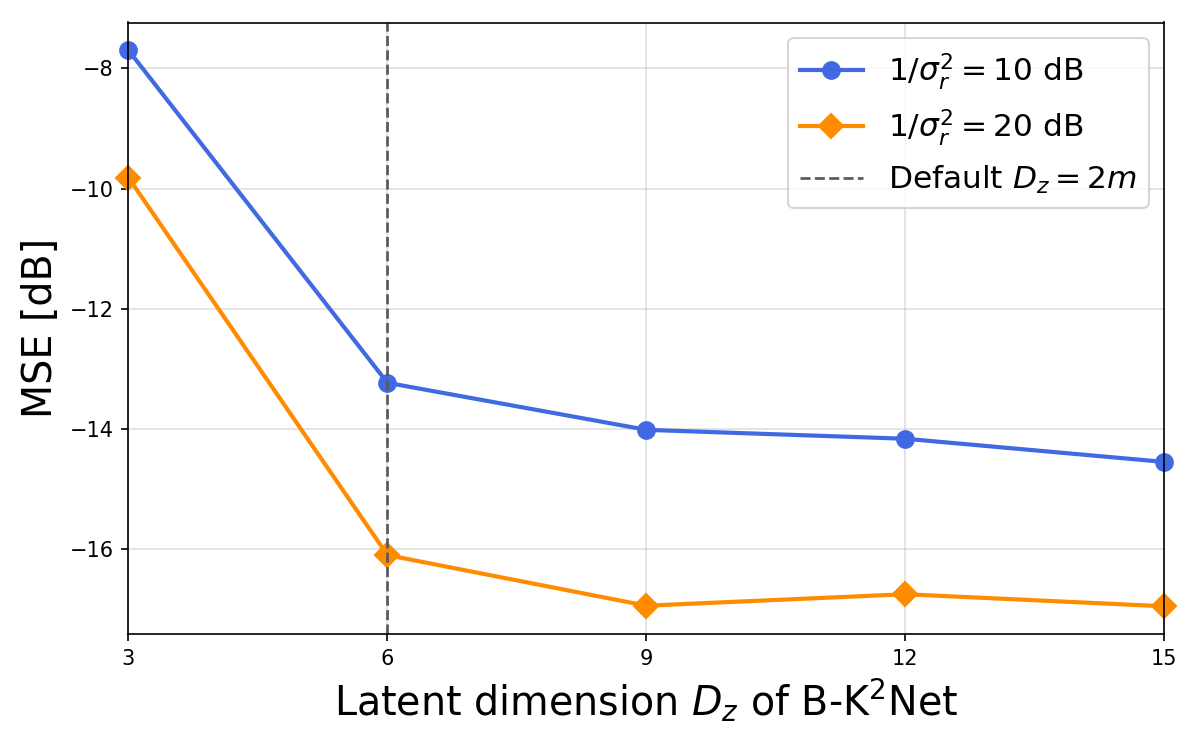}}}
    \vspace{-0.2cm}
    \caption{B-K$^2$Net performance versus the latent dimension $D_z$.}
    \label{fig:Dz_sensitivity}} 
\end{figure}

\begin{figure*}[!t]
    \centering
    $\begin{array}{c}
    {\resizebox{0.449\textwidth}{!}{\includegraphics{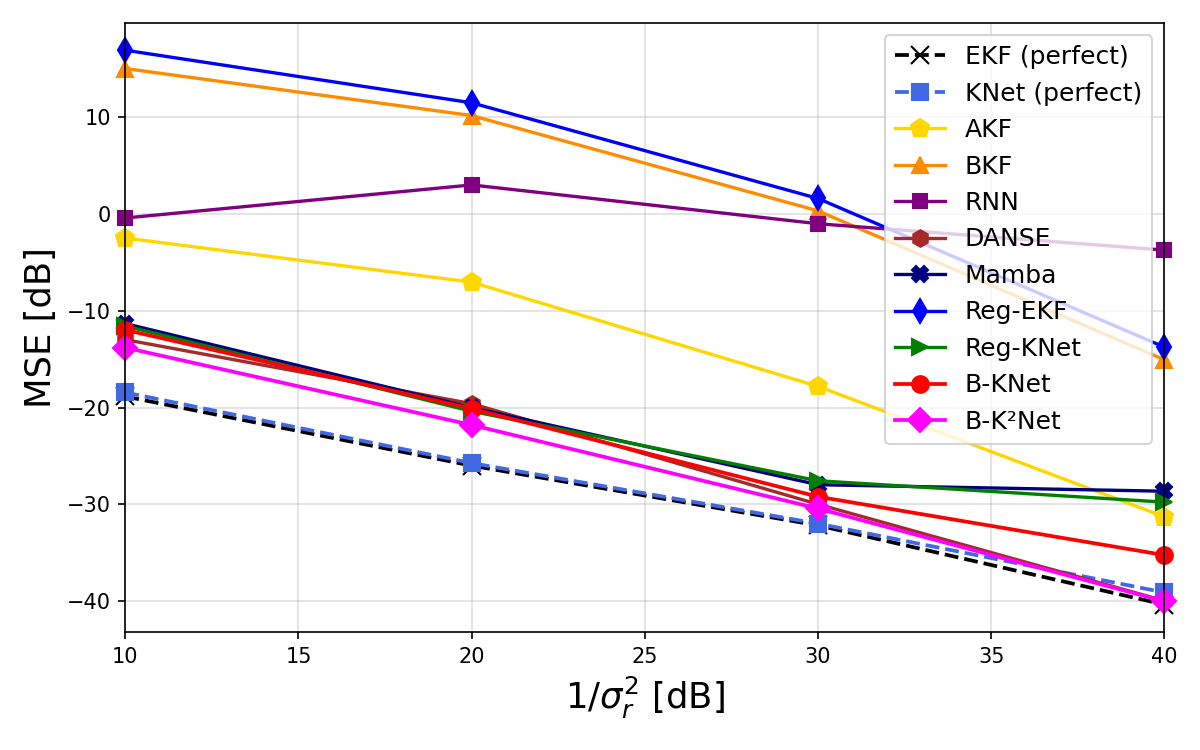}}
    } 
    \\ 
    \mbox{\small (a) $\bh(\cdot)$: linear}
    \end{array}
    $
    \hspace{0.01\textwidth}
    $\begin{array}{c}
    {\resizebox{0.449\textwidth}{!}{\includegraphics{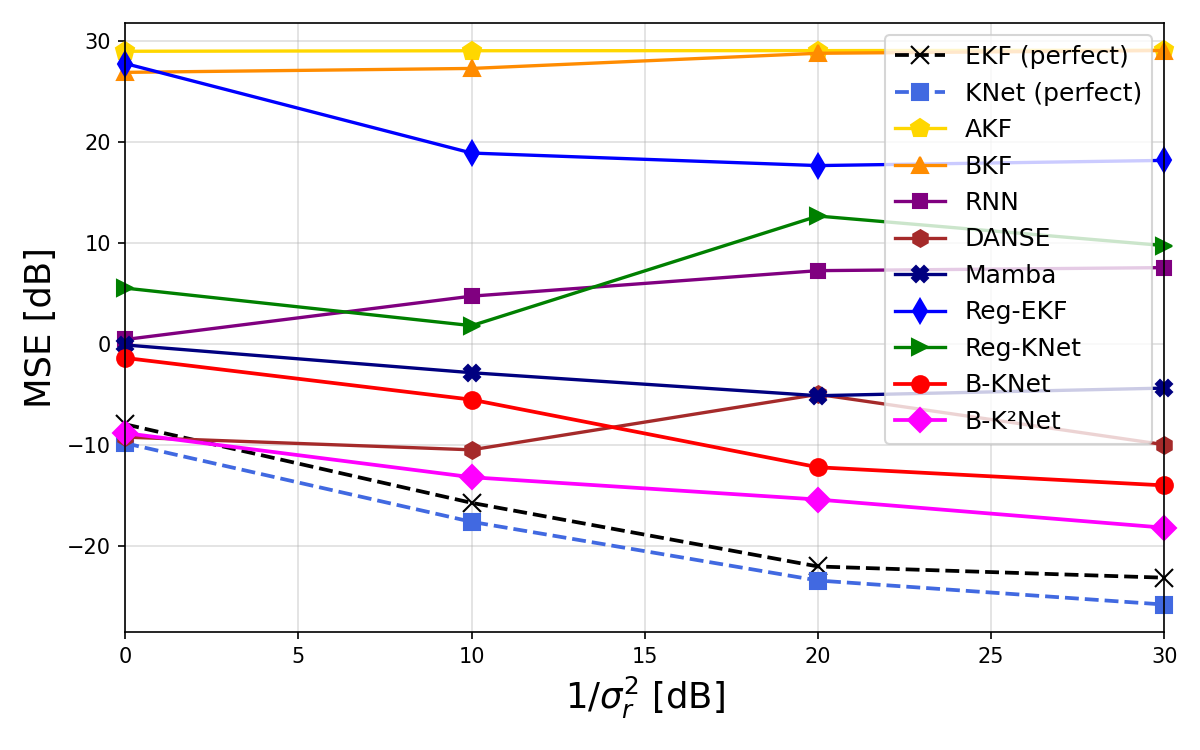}}
    }
    \\  
    \vspace{-0.1cm}
    \mbox{\small (b) $\bh(\cdot)$: nonlinear}
    \end{array}$
    \caption{The MSE results of Lorenz attractor with linear and nonlinear observation models.}
    \label{fig:Lorenz}
\end{figure*}

\begin{figure*}[!t]
    \centering
    \includegraphics[width=1\textwidth]{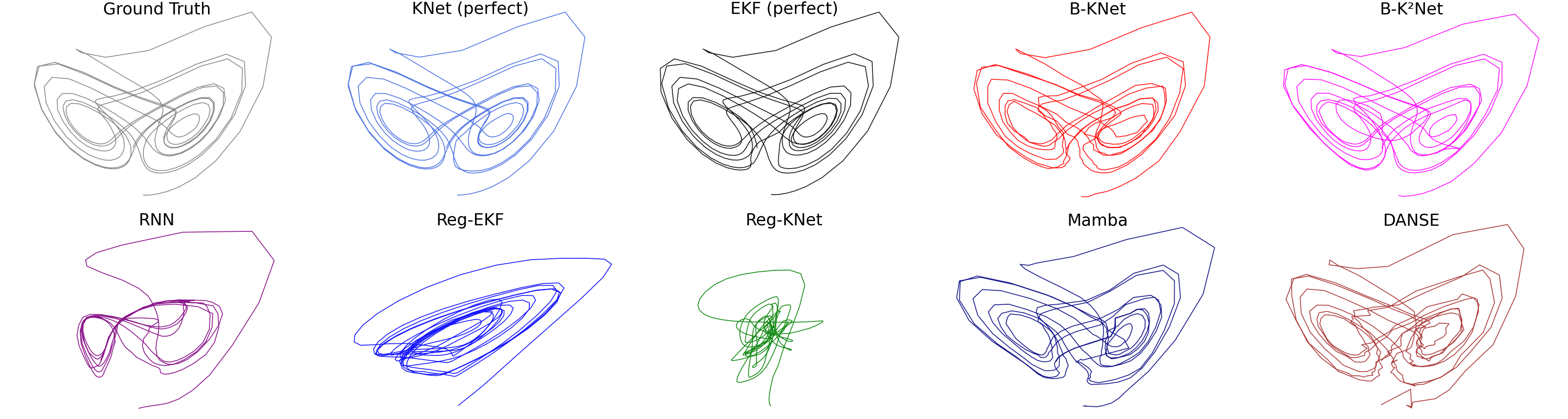}
    \captionsetup{justification=centering}
    \vspace{-0.5cm}
    \caption{Estimated trajectories on Lorenz attractor with a nonlinear observation model, $T_{\rm train}=150$ and $T_{\rm test}=400$.}
    \label{fig:Lorenz_traj}
\end{figure*}

We consider the Lorenz attractor as a canonical chaotic nonlinear dynamics with $m=n=3$, following the SS model formulation in \cite{revach2022kalmannet}. 
The continuous-time process $\bx_{\tau}$ evolves:
\begin{align}
    \label{eq:Lorenz}
    \frac{\partial \bx_{\tau}}{\partial \tau} = \bA (\bx_{\tau})\, \bx_{\tau},\;\; \bA(\bx_{\tau}) = 
    \begin{pmatrix}
        -10 & 10 & 0
        \\
        28 & -1 & -x_{1, \tau}
        \\
        0 & x_{1, \tau} & -\frac{8}{3}
    \end{pmatrix},
\end{align}
whose entries follow the standard Lorenz parameters.
Discretizing with a fixed step size $\Delta\tau$ and assuming
$\bA(\bx_\tau)$ is approximately constant over $[\tau,\tau+\Delta\tau]$, the transition matrix is obtained by truncating the matrix exponential at order $J$ as
\begin{align}
    \label{eq:Loren_dis}
    \bF(\bx_{\tau}) \triangleq e^{\bA(\bx_\tau)\Delta\tau}
    \approx \bI + \sum\nolimits_{j=1}^{J} \frac{(\bA(\bx_{\tau})\Delta\tau)^{j}}{j!}.
\end{align}
The resulting discrete-time SS model with additive white Gaussian process noise is then given by
\begin{align}
    \label{eq:Lorenz_ss}
    \bx_t = \bF(\bx_{t-1})\, \bx_{t-1} + \be_t, \quad \bQ = \sigma_{\sf q}^2 \bI.
\end{align}

For the Lorenz system, we consider linear/nonlinear observation models as follows:
\begin{align}
    \by_t =
    \begin{cases}
        \bx_t + \bn_t, & \bh(\cdot)\!\!:\text{linear},\\[2mm]
        \left[r_t, \varphi_t, \vartheta_t \right]^{\sf T} + \bn_t,
        & \bh(\cdot)\!\!:\text{nonlinear},
    \end{cases}
\end{align}
where $r_t$ is the range, $\varphi_t$ is the azimuth, and $\vartheta_t$ is the polar angle from $\bx_t$, which maps cartesian states to spherical coordinates as considered in \cite{revach2022kalmannet}.
For nonlinear observation, we use independent Gaussian observation noise with
\begin{align}
    \bR = \mathrm{diag}\big(\sigma_{\rm r}^2,\sigma_\varphi^2,\sigma_\vartheta^2\big).
\end{align}

In the simulations, we set $J=5$ Taylor order, $\Delta\tau = 0.03$ sampling interval, and $\sigma^2_{\rm q} = 10^{-3}$.
We use $T_{\rm train}=60$ and $T_{\rm test}=2000$ for the linear observation model, and use $T_{\rm train}=50$ and $T_{\rm test}=50$ for the nonlinear observation model.

To examine the sensitivity to the latent dimension, Fig.~\ref{fig:Dz_sensitivity} compares B-K$^2$Net for $D_z\in\{m,2m,3m,4m,5m\}$.
Increasing $D_z$ from $m$ to $2m$ substantially improves the estimation performance, demonstrating the benefit of augmenting the state with nonlinear systems.
Beyond $D_z=2m$, the performance varies only modestly at both noise levels. 
Accordingly, we adopt $D_z=2m$ throughout the experiments as a favorable performance--complexity trade-off.

Fig.~\ref{fig:Lorenz} shows the MSE versus $1/\sigma_{\rm r}^2$ under the
linear and nonlinear observation models.
In Fig.~\ref{fig:Lorenz}(a), under linear observation, the model-based
baselines without access to the true dynamics struggle to track the chaotic
Lorenz dynamics using a single linear surrogate.
Reg-KNet substantially improves over Reg-EKF by learning the correction gain
from data, but it still relies on the fixed $\widehat{\bF}$ and thus
saturates at low observation-noise levels, where a single linear surrogate
cannot adequately capture the state-dependent nonlinear evolution.
In contrast, the proposed B-KNet and B-K$^2$Net continue to improve as
$1/\sigma_{\rm r}^2$ increases and approach the oracle KNet (perfect), with
B-K$^2$Net providing an additional gain through its structured
Koopman predictor and task-aware residual adaptation.
DANSE also achieves performance comparable to B-K$^2$Net in this regime by
learning the predictive state distribution directly from the measurement
history without requiring an explicit state-transition model.

The advantage of DANSE under linear observations is reduced when the
observation model becomes nonlinear.
In nonlinear observations, DANSE now relies on a particle-based approximation of the posterior.
Under this more challenging setting, as shown in Fig.~\ref{fig:Lorenz}(b), the non-oracle AI-aided baselines exhibit a persistent performance gap relative to the proposed methods across the considered noise levels.
The results highlight the benefit of combining a structured predictor with
task-aware correction when the dynamics is unknown and the
observation model is nonlinear.

Fig.~\ref{fig:Lorenz_traj} shows the estimated trajectories with
$T_{\rm train}=150$ and $T_{\rm test}=400$ under the nonlinear observation
model at $1/\sigma_{\rm r}^2 = 20$ dB.
While Reg-KNet fails to recover the chaotic Lorenz dynamics, Mamba and DANSE
capture the overall two-lobe structure of the attractor but exhibit more
irregular and distorted trajectories.
In comparison, the proposed B-KNet and B-K$^2$Net preserve both the global
attractor structure and the local trajectory evolution more faithfully.
\\

\subsection{Real-World Dynamics} \label{subsec:realworld}
\begin{table}[!t]
\centering
\caption{Numerical Results for the UTIL Experiment with $T_{\rm test}=100$ and $N_T = 3$.}
\label{tab:UTIL}
\setlength{\tabcolsep}{8pt}
\vspace{-0.5em}
\renewcommand{\arraystretch}{1.15}
\begin{tabular}{l ccc}
\toprule
Method & RMSE [m] & MSE [dB] & Run time [sec] \\
\midrule
EKF \cite{zhao2024util}
& $0.1549$
& $-17.690$
& $0.0711$ \\

AKF
& $1.4718$
& $4.493$
& $0.3284$ \\

BKF
& $1.1642$
& $2.955$
& $1.2147$ \\

Mamba
& $0.1623$
& $-17.124$
& $0.0814$ \\

DANSE
& $0.1616$
& $-16.222$
& $0.0766$ \\

Reg-EKF
& $0.1466$
& $-17.928$
& $\mathbf{0.0681}$ \\

Reg-KNet
& $0.1324$
& $-18.631$
& $0.0942$ \\

\midrule
\textbf{B-KNet}
& $0.1172$
& $-19.626$
& $0.1696$ \\

\textbf{B-K$^{\mathbf{2}}$Net}
& $\mathbf{0.1113}$
& $\mathbf{-20.597}$
& $0.1849$ \\

\bottomrule
\end{tabular}
\end{table}
\begin{figure}[!t]    
    {\centerline{\resizebox{0.88\columnwidth}{!}{\includegraphics{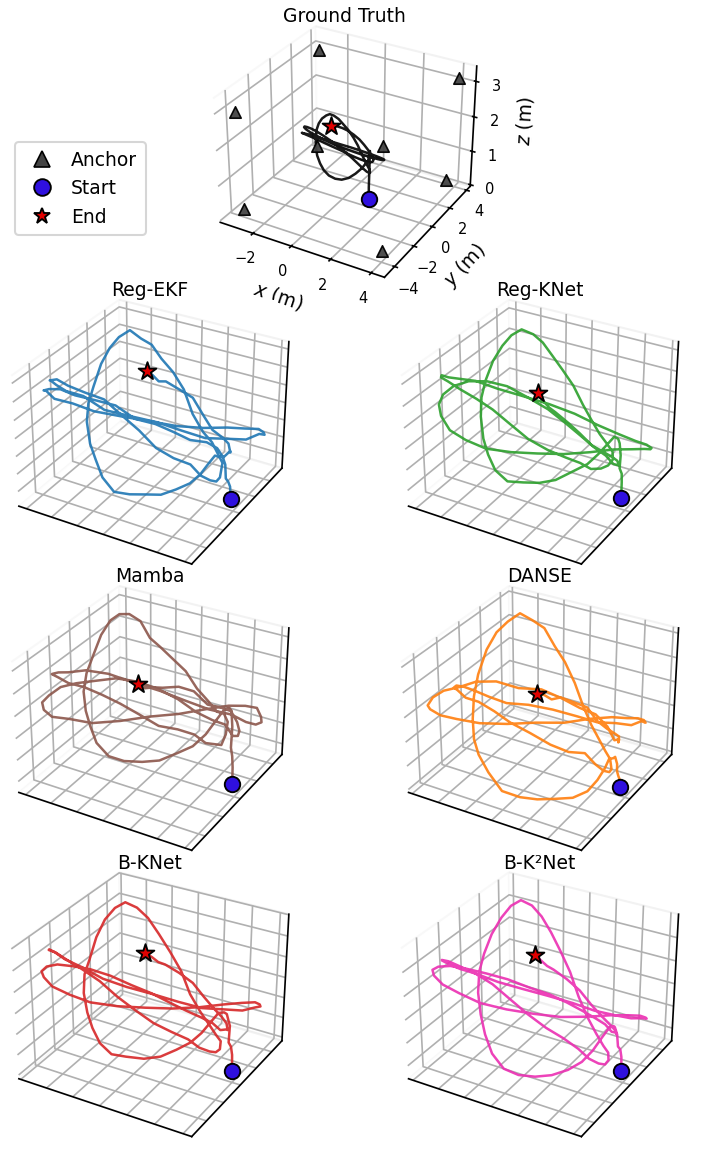}}}
    \vspace{-0.2cm}
    \caption{UTIL dataset: ground truth vs. estimated trajectories.}
    \label{fig:UTIL_traj}} 
\end{figure}

To evaluate the proposed frameworks under real-world dynamics
with non-Gaussian noise arising from practical measurement non-idealities, we use the UTIL dataset~\cite{zhao2024util}, which provides labeled drone flight trajectories collected with an ultra-wideband indoor localization system.
The dataset provides $20$ flight trials across three line-of-sight (LoS) anchor constellations (\texttt{const1}\,–\,\texttt{const3}), each recording $2000$ samples ($100$\,sec at $20$\,Hz) of $n=8$ time-difference-of-arrival (TDOA) measurements together with millimeter-accurate Vicon ground truth.  
We adopt a constant-velocity state $\bx_t=\big[\bp_t^{\sf T},\bv_t^{\sf T}\big]^{\sf T}\!\in\!\mathbb{R}^{6}$ where $\bp_t\in \bbR^3$ and $\bv_t\in \bbR^3$ denote the 3D position and velocity at time $t$, respectively, and the nonlinear  observation model as 
\begin{align}
    \bh_j(\bx_t)\!=\!\|\bp_t-\ba_{j_1}\| -\|\bp_t-\ba_{j_0}\|, 
    \quad j=1,\dots,n,
    \label{eq:UTIL_h}
\end{align}
where $\{\ba_{j_0},\ba_{j_1}\}$ are the two anchors defining the $j$-th TDOA pair.
Note that $\bh(\cdot)$ is anchor-set-dependent and varies trial-to-trial.
For the model-based baselines that require explicit noise statistics, $\bR=\sigma_{\rm r}^2\bI$ is empirically calibrated from the per-pair TDOA residuals measured on the dataset ($\sigma_{\rm r}\!\approx\!0.21$ m), and $\bQ$ is block-diagonal with hand-tuned position and velocity entries selected on the validation trials.
The proposed B-KNet and B-${\rm K}^{2}$Net, in contrast, do not require 
$\bQ$ or $\bR$: their gains are learned end-to-end from data, which is 
one of the practical advantages of the unknown-dynamics setting.

We split the $20$ LoS trials in a stratified manner and downsample each by a factor of $20$ to an effective rate of $1$\,Hz, yielding trials of length $100$.
Each training and validation trial is further split into non-overlapping segments of length $T_{\rm train}=30$, producing $N_E = 39$ and $N_V = 9$ datasets, while the $N_T = 3$ test trials are used in full with $T_{\rm test}=100$.

We evaluate position-estimation accuracy by the empirical root-mean-square error (RMSE) on the position sub-state $\bp_t$: $\mathrm{RMSE}\,{=}\,\sqrt{\frac{1}{T}\sum_{t=1}^{T}\left\|\hat{\bp}_t - \bp_t\right\|^2}$
in meters with the estimated position $\hat{\bp}_t$.
We also include the default EKF baseline in \cite{zhao2024util}, denoted EKF~\cite{zhao2024util}, which uses hand-tuned $\bQ$ and $\bR$ matched to the dataset.

Table~\ref{tab:UTIL} shows the mean $3$D position RMSE, MSE, and 
the per-trajectory run time on the test trials~\cite{zhao2024util}. 
Classical baselines fail to track the drone, returning errors of order one meter.
Among the model-based filters, the regression-based EKF and EKF~\cite{zhao2024util} both attain RMSE around $15$\,cm, consistent with the per-pair empirical TDOA noise floor of $\sigma_{\rm r}\!\approx\!0.21$\,m measured in the dataset.
KalmanNet variants significantly outperform these model-based filters while maintaining comparable run times, owing to the simple GRU-based architecture inherited from KalmanNet. 
In particular, although B-${\rm K}^{2}$Net additionally employs the DKN, its lightweight feedforward design keeps the run time comparable to Reg-KNet.
Among the KalmanNet variants, B-KNet further reduces the tracking error, and B-K$^2$Net attains the lowest MSE and RMSE among all compared schemes.
Notably, these gains are achieved using only 39 training segments, demonstrating effective learning with a relatively small dataset in this real-world setting.
Moreover, B-${\rm K}^{2}$Net achieves this superior accuracy with a moderate increase in run time compared to Reg-KNet and DANSE.

Estimated trajectories for one of the test trials are shown in Fig.~\ref{fig:UTIL_traj}. 
Our proposed methods track the ground-truth flight pattern with less jitter than Reg-EKF and stay closer to the reference trajectory while finishing nearer the true endpoint.
Overall, these results confirm the practical effectiveness of the proposed frameworks under real-world dynamics with non-idealities in the unknown-dynamics systems.

Comparing the two proposed frameworks, B-K$^2$Net consistently attains the best accuracy at the cost of a two-stage training pipeline and more parameters than B-KNet.
B-KNet is thus preferable under tight training or complexity budgets and when the dynamics is close to linear, as in the UCM example, whereas B-K$^2$Net is the method of choice for strongly nonlinear dynamics.

\section{Conclusion} \label{sec:conclusion}

In this paper, we proposed task-aware neural Kalman filtering frameworks for sequential state estimation when both the state-evolution function and the noise statistics are unknown. 
To this end, we formulated a task-aware learning principle in which every learned component, the predictor as well as the correction gain, is trained jointly under the posterior state-estimation objective. 
Blind-KalmanNet then realizes this principle in its most direct form, carrying the learning principle of the Kalman gain into the prediction step by jointly learning a state-dependent linear surrogate and the Kalman gain. 
Building on this design, Koopman-aided Blind-KalmanNet, the main framework of this paper, lifts the unknown dynamics into a latent space where the state evolves linearly, a representation naturally favorable for Kalman filtering; the pre-trained Koopman predictor serves as a globally structured backbone, augmented by a task-aware residual surrogate and a learned Kalman gain inherited from Blind-KalmanNet. 
Experiments on synthetic and real-world dynamics show that both frameworks remain competitive with existing AI-aided baselines. Koopman-aided Blind-KalmanNet attains the best accuracy across all considered settings, whereas Blind-KalmanNet recovers most of the achievable gain when the dynamics is close to linear, at a smaller model size and with a single-stage training pipeline.
Moreover, these results are achieved using a relatively small dataset, while the model size and the inference complexity remain comparable to those of the original KalmanNet.

\bibliographystyle{IEEEtran}
\bibliography{ref}

\end{document}